\documentclass[aps,pra,reprint,nofootinbib,superscriptaddress,twocolumn,showpacs,showkeys,longbibliography,amsmath,amssymb]{revtex4-2}
\usepackage{graphicx}
\usepackage{dcolumn}
\usepackage{bm}
\usepackage{braket}
\usepackage{subfigure}
\usepackage[colorlinks,bookmarks=false,citecolor=blue,linkcolor=red,urlcolor=blue]{hyperref}
\usepackage[english]{babel}
\usepackage{changes}
\usepackage{multirow}
\usepackage{booktabs}
\usepackage{float}
\usepackage{array} 
\begin{document}
\preprint{APS/123-QED}	
	
\title{Tuning spin-density separation via finite-range interactions: Dimensionality-driven signatures in dynamic structure factors}
\author{Xiaoran Ye}
\affiliation{Department of Physics, Zhejiang Normal University, Jinhua 321004, People's Republic of China}
\author{Yi Zhang}
\affiliation{Department of Physics, Zhejiang Normal University, Jinhua 321004, People's Republic of China}
\author{Ziheng Zhou}
\affiliation{Department of Physics, Zhejiang Normal University, Jinhua 321004, People's Republic of China}
\author{Zhaoxin Liang}\email[Corresponding author:~] {zhxliang@zjnu.edu.cn}
\affiliation{Department of Physics, Zhejiang Normal University, Jinhua 321004, People's Republic of China}

\date{\today}

\begin{abstract}
Spin-density (charge) separation, marked by distinct propagation velocities of spin and density excitations, epitomizes strong correlations, historically confined to one-dimensional (1D) systems. The recent experimental work of \href{https://doi.org/10.1038/s41586-025-09016-9}{S. Dhar, B. Wang, M.  Horvath, {\it et al.}  Nature {\bf 642}, 53 (2025)}, using a weakly interacting 3D Bose-Einstein condensate of $^{133}$Cs atoms confined in a 2D optical lattice to realize spin-density separation and demonstrate boson anyonization, motivates a deeper exploration into how dimensionality and interactions govern quantum correlations.  In this work, we investigate this in two-component bosonic mixtures with finite-range interactions, probing 1D and 3D dynamics. Using path integral effective field theory within the one-loop approximation, we derive analytical expressions for zero-temperature ground-state energy and quantum depletion, seamlessly recovering contact interaction results in the contact limit. By crafting an effective action for decoupled density and spin modes, we compute dynamic structure factors (DSFs), revealing how finite-range interactions sculpt spin-density separation. A pivotal finding is the dimensionality-driven divergence in DSF peak dynamics: in 1D, peaks ascend to higher frequencies with increasing interaction strength, yielding sharp responses; in 3D, peaks descend to lower frequencies, with broader density wave profiles. These insights highlight dimensionality's critical role in collective excitations and provide a robust theoretical blueprint for probing interaction-driven quantum phenomena via Bragg spectroscopy, paving new pathways for exploring dimensionally tuned quantum correlations in ultracold quantum gases.
\end{abstract}
\keywords{Ultracold quantum gas, Effective field theory, Dynamical structure factor, Quantum depletion, Spin-density (charge) separation}
\maketitle

\section{INTRODUCTION}

Spin-density separation, a defining feature of one-dimensional (1D) quantum many-body systems, emerges as the decoupling of low-energy collective excitations into independent spin and density modes, elegantly captured by the Tomonaga-Luttinger liquid framework for 1D Fermi gases~\cite{Tomonaga1950,Luttinger1963,Haldane1981,Volt1993,Cheianov2004}. Governed by tunable interactions, this phenomenon results in distinct linear dispersions for spin and density excitations, with velocities determined by Luttinger parameters and interaction strength~\cite{Feng2020,Recati2003,Scopa2021,Lee2012,Zacher1998,Polini2007}. Experimental techniques, such as Bragg spectroscopy~\cite{Ruwan2022,Yang2018} and tunneling spectroscopy~\cite{Jompol2009,Auslaender2005}, probe the dynamic structure factor (DSF) and tunneling conductance of one-dimensional quantum systems, revealing distinct spectral features of spin and density excitations, which align with theoretical predictions and highlight the role of interaction strength in shaping excitation dynamics~\cite{Kim1996}. Although primarily observed in 1D Fermi systems, spin-density separation extends to 1D Bosonic mixtures, suggesting its potential in higher-dimensional bosonic systems~\cite{Kleine2008}. In a 1D bosonic gas, spin-charge separation enables the experimental observation of anyonic correlations, highlighting tunable statistical phases via strong interactions~\cite{Dhar2025}. However, whether spin-density separation persists in higher-dimensional systems remains an important question, reflecting the critical role of dimensionality in modulating quantum correlations and collective phenomena.

Along this research line, spin-density separation in two-component Bosonic mixtures in higher dimensions is confirmed to exist without relying on bosonization and is driven by distinct interaction types. In higher-dimensional Bosonic mixtures, weakly interacting contact interactions govern the separation~\cite{Chung2008}. Meanwhile, $P$-wave interactions in three-dimensional (3D) bosonic mixtures modulate the separation dynamics, shaping a single degree-of-freedom peak observable in Bragg spectroscopy measurements~\cite{Ye2025}. Nevertheless, how interactions with more complex degrees of freedom—such as finite-range interactions—enable precise control of spin-density separation remains an important question. This reflects the critical role of complex interatomic interactions in shaping quantum correlations and collective phenomena, particularly through their influence on excitation spectra and mode decoupling mechanisms.

Another important motivation for this work stems from the emergence of exotic many-body phenomena---including quantum droplets~\cite{Petrov2016,Hu2020,Petrov2015}, phase separation~\cite{Timmermans1998,Buchler2003,Navarro2009}, mixed bubble states~\cite{Naidon2021,Sturmer2022}, and miscible-immiscible transitions~\cite{Wen2012,Thalhammer2008,Wen2020}---in two-component Bosonic mixtures. These phenomena are driven by the intricate interplay between interactions and quantum fluctuations. In the weakly interacting regime, they are predominantly governed by the competition between intra- and interspecies $s$-wave contact interactions, where the $s$-wave scattering length serves as the primary tunable parameter~\cite{Papp2008,Roy2015,Petrov2018}. Beyond this standard paradigm, a diverse array of interactions---such as dipole-dipole interactions, Rabi coupling, spin-orbit coupling, and finite-range interactions---introduce additional degrees of freedom. These include non-locality, spin coherence, and momentum-dependent coupling, enabling precise sculpting of the effective interatomic potential~\cite{Bisset2021,Cappellaro2017,Cui2018,Li2019,Boudjemaa2018,Suzuki2008,Gautam2014,Chiquillo2019}. Such interactions significantly expand the spectrum of accessible quantum phases. Crucially, finite-range interactions have recently emerged as a critically important control parameter~\cite{Cappe2017,Tononi2018,Zhang2024,Ye2024,Salasnich2017,Yu2024}. In two-component systems, they play a pivotal role in stabilizing quantum droplets and modulating density profiles beyond the contact-interaction approximation~\cite{Chiquillo2025,Emerson2025}. Furthermore, incorporating intra- and interspecies effective ranges ($r_{\text{e}}$, $r_{\text{e}_{12}}$) facilitates independent control over both density and spin channels, enabling comprehensive interaction-driven modulation of collective excitations.

In this work, we employ effective field theory within the one-loop approximation to derive analytical expressions for a bosonic mixture subjected to finite-range interactions at zero temperature. For both 3D and 1D systems, we compute the ground-state energy, which recovers the results for homogeneous Bosonic mixtures reported in Ref.~\cite{Chiquillo2025}, while the quantum depletion aligns with findings in Refs.~\cite{Chiquillo2018,Cappe2017} in the limit of vanishing finite-range interactions. A primary focus of this work is the formulation of an effective action that governs the decoupled density and spin degrees of freedom, enabling the calculation of the corresponding dynamic structure factors (DSFs). These reveal how finite-range interactions drive spin-density separation, a phenomenon amenable to experimental detection. Notably, this separation manifests distinct dimensional dependencies, with DSF peaks shifting to higher frequencies in 1D and lower frequencies in 3D as interaction strength increases, underscoring the pivotal role of dimensionality. In the absence of finite-range interactions, our results converge with those of Ref.~\cite{Chung2008}. These findings provide deep insights into the influence of finite-range interactions on spin-density separation in quantum many-body systems across varying dimensions.

The paper is organized as follows. In Sec.~\ref{2}, we develop a model for a two-component Bosonic mixtures in both the continuous 1D limit and 3D scenarios under finite-range interactions, utilizing an effective field theory approach. In Sec.~\ref{3}, we derive analytical expressions for the ground-state energy and quantum depletion under finite-range interactions within the same theoretical framework. In Sec.~\ref{4}, we explore the impact of finite-range interactions on spin-density separation across dimensions, examining the distinct behaviors in 1D and 3D systems through DSFs. Sec.~\ref{5} summarizes our findings and outlines the experimental conditions necessary for observing the proposed phenomena.

\section{FUNCTIONAL INTEGRATION OF THE $D$-DIMENSIONAL  Bosonic MIXTURES\label{2}}

\subsection{Partition function of the system}\label{PartitionS}

In this work, we are interested in a two-component weakly interacting Bosonic mixtures with finite-range interatomic interactions~\cite{Chiquillo2025} in spatial dimensions ($d=1,3$), confined to a volume $L^d$. To analyze this system, we employ the path-integral formalism~\cite{Ye2024,Yu2024,Ye2025,Zhang2024}, which naturally incorporates quantum fluctuations and enables systematic treatment of finite-range effects. The grand-canonical partition function of the system is then expressed as
\begin{eqnarray}
	\label{partition function}
	\mathcal{Z} & = & \int\mathcal{D}\left[\bar{\Psi},\Psi\right]\exp\left\{-\frac{S\left[\bar{\Psi},\Psi\right]}{\hbar}\right\},
\end{eqnarray}
with the action functional in Eq.~(\ref{partition function}) reading  $S\left[\bar{\Psi},\Psi\right]=\int_{0}^{\hbar\beta}\mathrm{d} \tau\int_{L^d}\mathrm{d}^{d}\mathbf {r} ~\mathcal{L}\left[\bar{\Psi},\Psi\right]$. The concrete expression of Lagrangian density $\mathcal{L}\left[\bar{\Psi},\Psi\right]$ here  takes the form  
\begin{eqnarray}	
\mathcal{L}&=&\sum_{i=1,2}\Bigg\{\psi_{i}^{*}\left[\hbar\partial_{\tau}-\frac{\hbar^{2}}{2m}\nabla^{2}-\mu\right]\psi_{i}\nonumber\\
&+&\frac{g^{\left(0\right)}}{2}\left|\psi_{i}\right|^{2}-\frac{g^{\left(2\right)}}{2}\left|\psi_{i}\right|^{2}\nabla^{2}\left|\psi_{i}\right|^{2}\Biggr\}
\nonumber\\
&+&g_{12}^{\left(0\right)}\left|\psi_{1}\right|^{2}\left|\psi_{2}\right|^{2}-g_{12}^{\left(2\right)}\left|\psi_{1}\right|^{2}\nabla^{2}\left|\psi_{2}\right|^{2}.\label{Ldensity}
\end{eqnarray}
In Eq.~(\ref{Ldensity}), the $\Psi\left(\boldsymbol{r}, \tau\right)=[\psi_1,\psi_2]^\mathrm{T}$ represent the two-component atomic bosonic complex fields, varying in both space $\boldsymbol{r}$ and imaginary time $\tau$. The parameters $\mu$ and $\beta = (k_{\text{B}} T)^{-1}$ respectively define the thermodynamic chemical potential and the inverse thermal energy scale governing system equilibrium. The $g^{(0)}$  and $g^{(0)}_{12}$ denote the intra- and interspecies  zero-range interatomic interaction coupling constants, respectively, with $g^{(0)}\neq g^{(0)}_{12}$ in view of the relevant experiment. Crucially, the higher-order terms $g^{(2)}$ and $g^{(2)}_{12}$ in Eq.~(\ref{Ldensity}) encode finite-range interaction effects through the $s$-wave pseudopotential expansion, where $\mathcal{O}(k^2)$ momentum dependence induces beyond-mean-field correlations. The explicit forms of the interaction parameters $g^{(0)}$ and $g^{(2)}$  will be systematically derived in the subsequent analysis, establishing their competition as the microscopic origin of spin-density separation in the phase diagram.

Our goal is to unravel the non-trivial interplay between dimensional confinement quantified by spatial dimensionality $d$ and the microscopic interaction structure encoded in contact $g^{(0)}$ and higher-order $g^{(2)}$ interaction parameters based on Lagrangian density $\mathcal{L}$ in Eq.~(\ref{Ldensity}), which orchestrates spin-density separation through emergent quantum fluctuations.

As a first step,  we derive the explicit functional dependence of the interaction parameters $g^{(0)}$ and $g^{(2)}$ on spatial dimensionality $d$ in Eq.~(\ref{Ldensity}). In 3D, the intra- and interspecies $s$-wave interaction strengths are defined as $ g^{\left(0\right)} = 4\pi \hbar^2 a / m $ and $ g^{\left(0\right)}_{12} = 4\pi \hbar^2 a_{12} / m $ respectively with $ a $ and $ a_{12} $ being the $s$-wave scattering lengths and $m$ being the atomic mass. The finite-range interaction strengths are $ g^{\left(2\right)}= 2\pi \hbar^2 a^2 r_{\text{e}} / m $ and $ g^{\left(2\right)}_{12} = 2\pi \hbar^2 a_{12}^2 r_{\text{e}_{12}} / m $ with $ r_{\text{e}} $ and $ r_{\text{e}_{12}} $ being the finite-range lengths.
 All discussions of 1D systems in this work refer to quasi-1D cases, realized through strong transverse confinement with frequency $ \omega_{\perp} $ and length $ l_{\perp} = \sqrt{\hbar / m \omega_{\perp}} $~\cite{Olshanii1998}. The effective $s$-wave interaction strengths are 
\begin{eqnarray}
	g^{\left(0\right)\left(\text{1D}\right)} &=& \frac{2 \hbar^2 a^{\left(\text{1D}\right)}}{m l_{\perp}^2} \frac{1}{1 - C a^{\left(\text{1D}\right)} / l_{\perp}},\nonumber\\ 
	g^{\left(0\right)\left(\text{1D}\right)}_{12} &=& \frac{2 \hbar^2 a_{12}^{\left(\text{1D}\right)}}{m l_{\perp}^2} \frac{1}{1 - C a_{12}^{\left(\text{1D}\right)} / l_{\perp}},\nonumber
\end{eqnarray}
where $ a^{\left(\text{1D}\right)} $ and $ a_{12}^{\left(\text{1D}\right)} $ are the effective 1D scattering lengths, and $ C = -\zeta(1/2) / \sqrt{2} \simeq 1.0326 $ with $ \zeta(x) $ being the Riemann zeta function~\cite{Pitaevskii2016}. For strong confinement ($ l_{\perp} \gg a^{\left(\text{1D}\right)}, a_{12}^{\left(\text{1D}\right)} $), these simplify to $	g^{\left(0\right)\left(\text{1D}\right)}  \approx 2 \hbar^2 a^{\left(\text{1D}\right)} / m l_{\perp}^2 $ and $ 	g^{\left(0\right)\left(\text{1D}\right)}_{12} \approx 2 \hbar^2 a_{12}^{\left(\text{1D}\right)} / m l_{\perp}^2 $~\cite{Bergeman2003}. The finite-range interaction strengths are assumed to be $ g^{\left(2\right)\left(\text{1D}\right)}= -\hbar^2 r_{\text{e}}^{\left(\text{1D}\right)} / m $ and $ g^{\left(2\right)\left(\text{1D}\right)}_{12} = -\hbar^2 r_{\text{e}_{12}}^{\left(\text{1D}\right)} / m $, where $ r_{\text{e}}^{\left(\text{1D}\right)} $ and $ r_{\text{e}_{12}}^{\left(\text{1D}\right)} $ are the intra- and interspecies $s$-wave effective 1D ranges.

Next, we plan to take an experimental viewpoint to reformulate the Lagrangian density $\mathcal{L}$ in Eq.~(\ref{Ldensity}) using density ($n_\rho$) and spin ($n_\sigma$) fluctuations. The quantum fluctuations of these quantities of both $n_{\rho }$  and $n_{\sigma}$ can be directly accessible in cold-atom experiments through Bragg spectroscopy measurements of the DSF. Defining
$n_\rho = (|\psi_1|^2 + |\psi_2|^2)/\sqrt{2}$ and 
$n_\sigma = (|\psi_1|^2 - |\psi_2|^2)/\sqrt{2} $, we express $\mathcal{L}$ in Eq.~(\ref{Ldensity}) into the following form
\begin{eqnarray}
	\mathcal{L}  &=&  \bar{\Psi}\left(\hbar\partial_{\tau}-\frac{\hbar^{2}}{2m}\nabla^{2}-\mu\right)\Psi
	+\frac{g^{\left(0\right)}_{\rho}}{2}n_{\rho}^{2}+\frac{g^{\left(0\right)}_{\sigma}}{2}n_{\sigma}^{2}\nonumber \\
	&   -&\frac{g^{\left(2\right)}_{\rho}}{2}n_{\rho}\nabla^{2}n_{\rho}-\frac{g^{\left(2\right)}_{\sigma}}{2}n_{\sigma}\nabla^{2}n_{\sigma}.\label{action}
\end{eqnarray}
In Eq.~(\ref{action}), the contact couplings $g^{(0)}_\rho = g^{(0)} + g^{(0)}_{12}$ and $g^{(0)}_\sigma = g^{(0)} - g^{(0)}_{12}$ characterize density and spin waves, respectively; while finite-range couplings $g^{(2)}_\rho = g^{(2)} + g^{(2)}_{12}$ and $g^{(2)}_\sigma = g^{(2)} - g^{(2)}_{12}$ govern their beyond-contact counterparts.

Now, we are ready to investigate the non-trivial interplay between dimensional confinement and the microscopic interaction structure by calculating both the ground-state energy, quantum depletion and the DSF.
In this context, we limit ourselves to the case that intraspecies interactions are all repulsive ($g^{\left(0\right)}>0$, $g^{\left(2\right)}>0$) and we work in the superfluid phase, where a U(1) gauge symmetry of each component is spontaneously broken~\cite{Cappellaro2017}. Under this assumption, the complex fields in Eq.~(\ref{action}) can be expanded as $\psi_{i}\left({\boldsymbol{r}},\tau\right)=\sqrt{n_{0}}+\eta_{i}\left({\boldsymbol{r}},\tau\right)$, where $n_0 = |\psi^{\left(0\right)}_i|^2$ represents the 3D condensate or 1D quasi-condensate density within the Bogoliubov approximation~\cite{Chiquillo2018,Pitaevskii2016}, while $\eta_{i}^{*}\left({\boldsymbol{r}},\tau\right)$ and $\eta_{i}\left({\boldsymbol{r}},\tau\right)$ denote the fluctuation fields above the condensate. The mean-field plus Gaussian approximation is obtained by expanding Eq.~(\ref{action}) up to second order in $\eta_{i}^{*}\left({\boldsymbol{r}},\tau\right)$ and $\eta_{i}\left({\boldsymbol{r}},\tau\right)$. 

Finally, the grand potential corresponding to our model system is given by
\begin{eqnarray}
	\Omega\left(\mu,\sqrt{n_{0}}\right) =  \Omega_{0}\left(\mu,\sqrt{n_{0}}\right)+\Omega_{g}\left(\mu,\sqrt{n_{0}}\right),\label{GrandPot}
\end{eqnarray}
with $\Omega_{0}/L^d=  -2\mu n_{0}+g^{\left(0\right)}n_{0}^{2}+g^{\left(0\right)}_{12}n_{0}^{2}$
being the mean-field grand-potential, while $\Omega_{g}$ takes into account Gaussian fluctuations. Requiring $\sqrt{n_{0}}$ to describe the condensate (3D BEC or quasi-1D), we ensure vanishing linear fluctuation terms in the action, confirming $\sqrt{n_{0}}$ minimizes the action. Minimizing $\Omega_{0}$ via $\partial\Omega_{0}/\partial\sqrt{n_{0}} = 0$ yields the chemical potential $\mu = ( g^{(0)} + g^{(0)}_{12} ) n_{0}$.

\begin{figure}[t] 
	\begin{centering} 
		\includegraphics[scale=0.8]{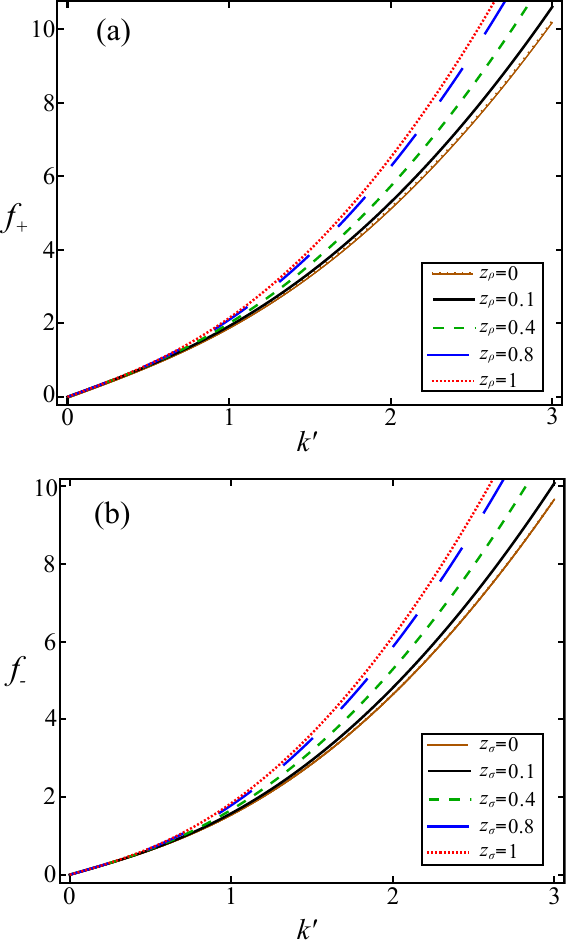} 
		\par\end{centering} 
	\caption{ (a) Dimensionless excitation spectrum $ f_{+}(k^{\prime}) $ from Eq.~\eqref{DEP} for varying $ z_{\rho} \equiv 4mn_{0}g_{\rho}^{(2)}/\hbar^{2} $.   (b) Corresponding spectrum $ f_{-}(k^{\prime}) $ for different $ z_{\sigma} \equiv 4mn_{0}g_{\sigma}^{(2)}/\hbar^{2} $.  The dimensionless interaction parameter is fixed at $ y = 0.3 $.
    \label{fig1}
}
\end{figure}

\subsection{Bogoliubov excitations} \label{S2B}

In the preceding subsection~\ref{PartitionS}, we have presented the system's action in Eq.~(\ref{Ldensity}), paying special attentions to  finite-range interaction effects in the two-component mixture characterized by the parameters of $g^{(2)}$. In this subsection~\ref{S2B}, we plan to derive the excitation spectrum from the Gaussian action as shown in Eq.~(\ref{GrandPot}). To diagonalize the quadratic terms in $\eta_{i}$, $\eta_{i}^{*}$ in Eq.~(\ref{action}), we Fourier transform and introduce Nambu space~\cite{Armaitis2015}, yielding the desired quadratic form
\begin{eqnarray}
	S_{2}\left[\eta_{1}^{*},\eta_{1},\eta_{2}^{*},\eta_{2}\right] & = & -\frac{\hbar}{2}\sum_{\boldsymbol{k}\ne0,n}{\bf \Phi}_{\boldsymbol{k}n}\boldsymbol{G}_{\boldsymbol{k}n}^{-1}{\bf \Phi}^{\dagger}_{\boldsymbol{k}n},\label{BogoS}
\end{eqnarray}
where $\hbar\boldsymbol{k}$ denotes the momentum and $n$ labels the Matsubara frequencies with $\omega_{n}=2\pi n/\hbar\beta$. The vector in Eq.~(\ref{BogoS})
\begin{eqnarray}
	{\bf \Phi}_{\boldsymbol{k}n} & = & \left(\eta_{1\boldsymbol{k}n}^{*},\eta_{1-\boldsymbol{k}n},\eta_{2\boldsymbol{k}n}^{*},\eta_{2-\boldsymbol{k}n}\right)\nonumber
\end{eqnarray}
resides in the appropriate Nambu space, and the inverse Green's function of the system in Eq. (\ref{BogoS}), $\boldsymbol{G}_{\boldsymbol{k}n}^{-1}$, can be written as
\begin{eqnarray}
	-\hbar\boldsymbol{G}_{\boldsymbol{k}n}^{-1} & = & \beta\left(\begin{array}{cc}
		-\hbar G_{B\boldsymbol{k}n}^{-1} & \hbar\Sigma_{12}\\
		\hbar\Sigma_{12} & -\hbar G_{B\boldsymbol{k}n}^{-1}
	\end{array}\right),\label{QGF}
\end{eqnarray}
with $G_{B\boldsymbol{k}n}^{-1}$ and $\Sigma_{12}$ being $2\times2$ submatrices. Henceforth, $2\times 2$ matrices are denoted by uppercase letters, while $4\times 4$ matrices are represented by bold uppercase letters. 

The diagonal submatrices in Eq.~(\ref{QGF}) are given by
\begin{eqnarray}
	-\hbar G_{B\boldsymbol{k}n}^{-1} & = & \left(\begin{array}{cc}
		-\mathrm{i}\hbar\omega_{n}+\epsilon^{0}_{k}+g^{\prime}n_{0}, & g^{\prime}n_{0},\\
		g^{\prime}n_{0}, & \mathrm{i}\hbar\omega_{n}+\epsilon^{0}_{k}+g^{\prime}n_{0},
	\end{array}\right),\nonumber \\
\end{eqnarray}
with $\epsilon^{0}_{k}=\hbar^{2}k^{2}/2m$ and $g^{\prime}=g^{\left(0\right)}+g^{\left(2\right)}k^{2}$. The off-diagonal matrix in Eq.~(\ref{QGF}), representing the self-energy due to interspecies coupling, is expressed as $\hbar\Sigma_{12}=\left(g^{\left(0\right)}_{12}+g^{\left(2\right)}_{12}k^{2}\right)n_{0}I$, where $I$ is a $2\times2$ matrix with all elements equal to $1$.

Furthermore, we define the matrix $\boldsymbol{M}$ as the inverse Green's function in Eq.~(\ref{QGF}) at $\omega_{n} = 0$:
\begin{eqnarray}
		\boldsymbol{M} \!\!& = \!\!& \left(\begin{array}{cccc}
			\!\!\epsilon^{0}_{k}+g^{\prime}n_{0}\! & g^{\prime}n_{0} & g^{\prime}_{12}n_{0} & g^{\prime}_{12}n_{0}\\
			g^{\prime}n_{0} & \!\!\epsilon^{0}_{k}+g^{\prime}n_{0}\!& g^{\prime}_{12}n_{0} & g^{\prime}_{12}n_{0}\\
			g^{\prime}_{12}n_{0} & g^{\prime}_{12}n_{0} & \!\!\epsilon^{0}_{k}+g^{\prime}n_{0}\!  & g^{\prime}n_{0} \\
			g^{\prime}_{12}n_{0} & g^{\prime}_{12}n_{0} & g^{\prime}n_{0}  & \!\!\epsilon^{0}_{k}+g^{\prime}n_{0}\! 
		\end{array}\right),\label{m}
	\end{eqnarray}
with $g^{\prime}_{12}=g^{\left(0\right)}_{12}+g^{\left(2\right)}_{12}k^{2}$. The Bogoliubov modes $E_{\pm}$ can be analytically obtained by letting $\det M=0$ in Eq.~(\ref{m}) with the help of the Cayley-Hamilton theorem (see Eq.~(\ref{AppE}) in Appendix~\ref{A} for the detailed derivation):
\begin{eqnarray}
	\!\!\!\!E_{\pm} \!= \! \sqrt{\frac{{\rm Tr}[\left(\boldsymbol{\kappa}\boldsymbol{M}\right)^{2}]}{4}\!\pm\!\sqrt{\frac{\left[{\rm Tr}\left(\boldsymbol{\kappa}\boldsymbol{M}\right)^{2}\right]^{2}}{16}\!-\!\det\left(\boldsymbol{\kappa}\boldsymbol{M}\right)}}. 
	\label{DE0}
\end{eqnarray}

The excitation energies of $E_{\pm}$ in Eq.~(\ref{DE0}) can be written into the dimensionless form as 
\begin{eqnarray}
	f_{\pm} = \frac{E_{\pm}}{g^{\left(0\right)}n_{0}} =\sqrt{\frac{\epsilon^{0}_{k}}{g^{\left(0\right)}n_{0}}\left(\frac{\epsilon^{0}_{k}}{g^{\left(0\right)}n_{0}}\lambda_{\pm}+2\pm2y\right)},\label{DEP}
\end{eqnarray}
where $y = g^{\left(0\right)}_{12}/g^{\left(0\right)}$ is the $s$-wave contact coupling ratio. The $f_\pm$ in Eq.~(\ref{DEP}) represent finite-range-interaction-modified Bogoliubov modes with $\lambda_\pm = 1 + z_{\rho(\sigma)}$. Here, the finite-range coupling strength $z_{\rho(\sigma)} = 4 m n_0 g^{\left(2\right)}_{\rho(\sigma)} / \hbar^2$ depends on the reduced couplings $g^{\left(2\right)}_{\rho(\sigma)} = g^{\left(2\right)}\pm g^{\left(2\right)}_{12}$ in the density ($\rho$) and spin ($\sigma$) channels. 

Figure~\ref{fig1} shows excitation spectra $f_\pm$ versus dimensionless wave vector $k^{\prime} = \hbar k / \sqrt{2mg^{(0)}n_0}$ for various finite-range couplings $z_{\rho(\sigma)}$. We have checked that Eq.~(\ref{DEP}) is simplified into Ref.~\cite{Abad2013} for vanishing the finite-range interaction coupling constant $z_{\rho(\sigma)}=0$. The competition between contact ($y$) and finite-range ($z$) interactions distinctly modifies density and spin excitations in Figs.~\ref{fig1}(a) and (b), respectively. In the phonon regime ($k \to 0$):
\begin{equation}
f_{\pm} \approx \sqrt{ \epsilon_k^0 (2 \pm 2y) / g^{(0)}n_0 } \nonumber
\end{equation}
showing negligible $z$-dependence as evident in Figs.~\ref{fig1} (a) and (b). Conversely, in the high-energy limit ($k \to \infty$):
\begin{equation}
f_{\pm} \approx \sqrt{\lambda_{\pm}} \, \epsilon_k^0 / g^{(0)}n_0, \nonumber
\end{equation}
where finite-range interactions ($z$) dramatically modify both fluctuations, evident in the $k^{\prime} \gg 0$ regimes.

\begin{table}[htbp]
    \renewcommand\arraystretch{1.5} 
    \setlength{\tabcolsep}{6pt} 
    \caption{Ground-state phase diagram of bosonic systems in the $y$-$z_{\rho(\sigma)}$ parameter space, where $y$ denotes the contact coupling ratio and $z_{\rho(\sigma)}$ the finite-range interaction strength (negative/positive values correspond to attraction/repulsion).}
    \label{table1}
    \begin{tabular}{>{\centering\arraybackslash}p{0.7cm}>{\centering\arraybackslash}p{1.8cm}>{\centering\arraybackslash}p{2.8cm}>{\centering\arraybackslash}p{1.2cm}}
        \toprule
        \textbf{Comp.} & \textbf{Parameters} & \textbf{Phase} & \textbf{Ref.} \\
        \midrule
        \multirow{3}{*}{Single} 
        & $y=0$, $z_{\rho(\sigma)}\ne0$ & Finite-range EOS & \cite{Cappe2017,Tononi2018,Salasnich2017,Zhang2024} \\
        & $y=0$, $z_{\rho(\sigma)}=0$ & Dimensional crossover & \cite{Orso2006,Hu2011,Zhou2010} \\
        & $y=0$, $z_{\rho(\sigma)}>0$ & Finite-range dimensional crossover & \cite{Ye2024,Yu2024} \\
        \midrule
        \multirow{6}{*}{Spinor}
        & $y \leq -1$, $z_{\rho(\sigma)} = 0$ & Quantum droplet & \cite{Petrov2015} \\
        & $y \leq -1$, $z_{\rho(\sigma)} > 0$ & Finite-range quantum droplet & \cite{Chiquillo2025,Emerson2025} \\
        & $0 < y < 1$, $z_{\rho(\sigma)}=0$ & Spin-density separation & \cite{Chung2008} \\
        & $0 < y < 1$, $0 \leq z_{\rho(\sigma)}\leq1$ & Spin-density separation (finite-range) & This work \\
        & $y>0$, $z_{\rho(\sigma)}=0$ & Mixed bubble phase & \cite{Naidon2021} \\
        & $y > 1$, $z_{\rho(\sigma)}=0$ & Phase separation & \cite{Alex2002} \\
        \bottomrule
    \end{tabular}
\end{table}

It is clear in Fig.~\ref{fig1} that the competition between the contact and finite-range interaction greatly affects  both the density and spin quantum fluctuations  of the model system, which directly determines the  ground-state phase diagram of model system. Hence, we are motivated to comprehensively summarize the ground-state phase diagram for bosonic systems across $y$ and $z_{\rho(\sigma)}$ in Table~\ref{table1}.

(i) Single-component system ($y = 0$): Finite-range interactions ($z_{\rho(\sigma)} \neq 0$) govern the Equation of state (EOS)~\cite{Cappe2017,Tononi2018,Salasnich2017,Zhang2024}. Positive $z_{\rho(\sigma)} > 0$ induces a dimensional crossover~\cite{Ye2024,Yu2024}, while $z_{\rho(\sigma)} = 0$ yields a distinct dimensional crossover~\cite{Orso2006,Hu2011,Zhou2010}.

(ii) Two-component system ($y \neq 0$):

    $z_{\rho(\sigma)} = 0$: Quantum droplet phase for $y \leq -1$ (stabilized by beyond-mean-field effects~\cite{Petrov2015}) and $-1 < y < 0$~\cite{Petrov2015}; spin-density separation for $0 < y < 1$~\cite{Chung2008}.
    
    $z_{\rho(\sigma)} \neq 0$: Quantum droplets persist for $y \leq -1$~\cite{Emerson2025,Chiquillo2025}. For $0 < y < 1$ and $0 \leq z_{\rho(\sigma)} \leq 1$, finite-range interactions modify spin-density separation (this work's focus). The system transitions to a mixed bubble phase at $y > 0$~\cite{Naidon2021} and phase separation at $y > 1$~\cite{Alex2002}.

We remark that our work together with the references in Table~\ref{table1} gives a complete description of how the contact and finite-range interaction couplings affecting  ground-state phase diagram for both single-component and two-component weakly-interacting Bosonic systems. 
\begin{figure}[hbtp] 
	\begin{centering} 
		\includegraphics[scale=0.8]{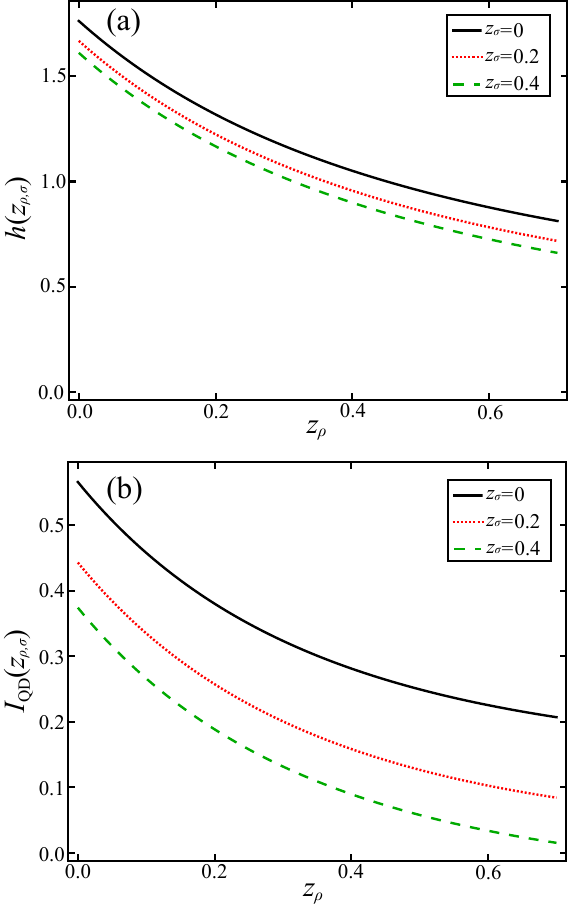} 
		\par\end{centering} 
\caption{(a) Scaling function $ h(z_{\rho}) $ in Eq.~\eqref{GEF} for $ z_{\rho} \equiv 4mn_{0} g_{\rho}^{(2)}/\hbar^{2} $ varied in $[0,0.7]$ and discrete values of $ z_{\sigma} \equiv 4mn_{0} g_{\sigma}^{(2)}/\hbar^{2} $. 
    (b) Scaling function $ I_\text{QD}(z_{\sigma}) $ in Eq.~\eqref{QDE} for $ z_{\sigma} \equiv 4mn_{0} g_{\sigma}^{(2)}/\hbar^{2} $ varied in $[0,0.7]$ and discrete values of $ z_{\rho} $.
    \label{fig2}
}
\end{figure}

\section{GROUND-STATE ENERGY AND QUANTUM DEPLETION IN $d$-DIMENSIONAL SYSTEM \label{3}}

In Sec.~\ref{2}, we have introduced the Lagrangian density of the system in Eq.~(\ref{Ldensity}) and derived the excitation spectrum in Eq.~(\ref{DE0}) within the framework of effective field theory. Next, in Sec.~\ref{3}, our goal is to derive explicit analytical expressions for the Lee-Huang-Yang (LHY) corrections to the ground-state energy and quantum depletion of $d$-dimensional Bosonic mixtures under the one-loop approximation. 

Our starting point is the grand potential of the Gaussian fluctuations $\Omega_{g}\left(\mu,\sqrt{n_{0}}\right)$ in Eq.~(\ref{GrandPot}), which is provided by
\begin{eqnarray}
	\Omega_{\text{g}}\left(\mu,\sqrt{n_{0}}\right)  =  \frac{1}{2}\sum_{\boldsymbol{k},\pm}\left[E_{\pm}+\frac{2}{\beta}\ln\left(1-\mathrm{e}^{-\beta E_{\pm}}\right)\right].
\end{eqnarray}
In what follows, our primary focus lies on the ground-state energy and quantum depletion of the model system at zero temperature.

\subsection{$d=3$}\label{31}

In this subsection~\ref{31}, we plan to investigate the ground-state energy and quantum depletion of the model system in 3D.  To this end, we compute the ground-state energy of the model system, $E_{\text{g}}=\Omega_{0}+\Omega_{\text{g}}+2L^{3}\mu n_{0}$.  Then, the expression for the ground-state energy can be written as follows:
	\begin{eqnarray}
		\!\!\!\!\!&&\frac{E_{\text{g}}}{L^{3}} =  g^{\left(0\right)}n_{0}^{2}+g^{\left(0\right)}_{12}n_{0}^{2}\nonumber\\
		\!\!\!\!\!&&+\frac{g^{\left(0\right)}n_{0}}{2L^{3}}\!\!\!\!\sum_{k^{\prime}\ne0,\pm}\!\!\left\{ f_{\pm}\!-\!\sqrt{\lambda_{\pm}}k^{\prime2}\!-\!\frac{1\pm y}{\sqrt{\lambda_{\pm}}}+\frac{\left(1\pm y\right)^{2}}{2\lambda_{\pm}^{\frac{3}{2}}k^{\prime2}}\right\} , 
		\label{GSE3}
	\end{eqnarray}
where the $k^{\prime}=\hbar k/\sqrt{2mg^{\left(0\right)}n_{0}}$ represents the dimensionless wave vector. 

In Equation~(\ref{GSE3}), the first and second term on the right side represent the mean-field contribution, while all the subsequent terms correspond to the beyond-mean field correction arising from quantum fluctuations. Notably, the final three terms in Eq.~(\ref{GSE3}) are incorporated to eliminate power-law ultraviolet divergences in the momentum summation, a procedure effectively subsumed within a suitable renormalization of the coupling constants~\cite{Braaten1997}. Rather than explicitly deriving the renormalized coupling constants, this subtraction yields a finite LHY correction, aligning with the physical goals of our study. The detailed derivation of these terms is presented in Appendix \ref{B}. In the continuum limit, the summation in Eq.~(\ref{GSE3}) can be systematically replaced by an integral, leading to the analytical expression for the ground-state energy corresponding to Eq. (\ref{partition function}) with $d=3$ as follows
\begin{equation}
	\frac{E_{\text{g}}}{L^{3}}  =  g^{\left(0\right)}n_{0}^{2}\!+\!g^{\left(0\right)}_{12}n_{0}^{2}\!+\! \frac{\left(g^{\left(0\right)}n_{0}\right)^{\frac{5}{2}}}{(2\pi)^{2}}\!\!\left(\frac{2m}{\hbar^{2}}\right)^{\frac{3}{2}}\!\!h_{\text{3D}}\left(z_{\rho,\sigma}\right),\label{GSE}
\end{equation}
where the scaling function $h_{\text{3D}}\left(z_{\rho,\sigma}\right)$ in terms of the variable $z_{\rho,\sigma}$ can be solved analytically and the result reads
\begin{equation}
	h\left(z_{\rho,\sigma}\right)  =\frac{8\sqrt{2}\left(1+y\right)^{\frac{5}{2}}}{15\left(1+z_{\rho}\right)^{2}}+  \frac{8\sqrt{2}\left(1-y\right)^{\frac{5}{2}}}{15\left(1+z_{\sigma}\right)^{2}}.\label{GEF}
\end{equation}
We remark that Eq.~(\ref{GSE}) is consistent with the corresponding one in Ref.~\cite{Chiquillo2025}. 

For the coefficient $y = 0.3$, the corresponding results of Eq.~(\ref{GEF}) are illustrated in Fig.~\ref{fig2}(a). In the absence of finite-range interactions, i.e., $z_{\rho(\sigma)} = 0$, the ground-state energy given by Eq.~(\ref{GSE3}) simplifies to
\begin{equation}
	\frac{E_{\text{g}}^{(z=0)}}{L^{3}} =  g^{\left(0\right)}n_{0}^{2}+g^{\left(0\right)}_{12}n_{0}^{2}+\frac{\left(g^{\left(0\right)}n_0\right)^{\frac{5}{2}}}{\left(2\pi\right)^{2}}\left(\frac{2m}{\hbar^{2}}\right)^{\frac{3}{2}}h^{\left(z=0\right)}\left(y\right), 
\end{equation}
where the function of $h_{\text{3D}}^{\left(z=0\right)}\left(y\right)$ is given by
\begin{equation}
	h^{\left(z=0\right)}\left(y\right) =  \frac{8\sqrt{2}}{15}\left(\left(1+y\right)^{\frac{5}{2}}+(1-y)^{\frac{5}{2}}\right),\label{GS0E}
\end{equation}
and the expression in Eq.~(\ref{GS0E}) accurately reproduces the relevant results in Ref.~\cite{Petrov2015}.

We  proceed to calculate the quantum depletion of our model system. The zero-temperature total particle number $N$ can be derived from the zero-temperature grand potential $\Omega_{0}+\Omega_{g}$ using the thermodynamic formula $N=-\partial\left(\Omega_{0}+\Omega_{g}\right)/\partial\mu$. Consequently, the quantum depletion $\Delta N=N-N_{0}$ of the system is obtained as
\begin{eqnarray}
\frac{\Delta N}{N} =  \frac{1}{(2\pi)^{2}}\left(\frac{2mg^{\left(0\right)}}{\hbar^{2}}\right)^{\frac{3}{2}}\left(g^{\left(0\right)}n\right)^{\frac{1}{2}}I_{\text{QD}}\left(z_{\rho,\sigma}\right),\label{QD}
\end{eqnarray}
where the scaling function $I_{\text{QD}}\left(z\right)$  is defined as
\begin{eqnarray}
I_{\text{QD}}\left(z_{\rho,\sigma}\right) = &-&\frac{\sqrt{2}\left(z_{\sigma}-1\right)\left(1-y\right)^{\frac{3}{2}}}{3\left(1+z_{\sigma}\right)^{2}}\nonumber \\
	& -&\frac{\sqrt{2}\left(y+1\right)^{\frac{3}{2}}}{3\left(1+z_{\rho}\right)}+\frac{2\sqrt{2}\left(y+1\right)^{\frac{3}{2}}}{3\left(1+z_{\rho}\right)^{2}}.\label{QDE}
\end{eqnarray}

For the parameter $y = 0.3$, the corresponding results of Eq.~(\ref{QDE}) are illustrated in Fig.~\ref{fig2}(b). To verify the correctness of these results, we consider the limit of vanishing finite-range interactions by setting $z_{\rho,\sigma} = 0$. In such a case, Eq.~(\ref{QD}) can be rewritten into
\begin{equation}
	\frac{\Delta N}{N}  =  \frac{1}{(2\pi)^{2}}\left(\frac{2mg^{\left(0\right)}}{\hbar^{2}}\right)^{\frac{3}{2}}\left(g^{\left(0\right)}n\right)^{\frac{1}{2}}I^{(0)}_{\text{QD}}\left(y\right),\label{QD0}
\end{equation}
with the scaling function $I^{(0)}_{\text{QD}}\left(y\right)$ reading
\begin{equation}
	I^{(0)}_{\text{QD}}\left(y\right)  =  \frac{\sqrt{2}}{3}\left[\left(1-y\right)^{\frac{3}{2}}+\left(y+1\right)^{\frac{3}{2}}\right].\label{QDE0}
\end{equation}
The analytical expression in Eq.~(\ref{QD0}) is fully consistent with the corresponding results in Ref.~\cite{Chiquillo2018}.

\subsection{$d=1$}

For a quasi-1D system, the ground-state energy corresponding to Eq.~(\ref{partition function}) with $d=1$ is expressed as:
\begin{widetext}
	\begin{eqnarray}
		\frac{E_{\text{g}}^{\left(\text{1D}\right)}}{L} & = & g^{\left(0\right)\left(\text{1D}\right)}n_{0}^{2}+g^{\left(0\right)\left(\text{1D}\right)}_{12}n_{0}^{2}+\frac{g^{\left(0\right)\left(\text{1D}\right)}n_{0}}{2L}\sum_{k^{\prime\left(\text{1D}\right)}\ne0,\pm}\left\{ f_{\pm}^{\left(\text{1D}\right)}-\sqrt{\lambda_{\pm}^{\left(\text{1D}\right)}}(k^{\prime\left(\text{1D}\right))})^2-\frac{1\pm y^{\left(\text{1D}\right)}}{\sqrt{\lambda_{\pm}^{\left(\text{1D}\right)}}}\right\} ,\nonumber \\
		\label{GSE1}
	\end{eqnarray}
\end{widetext}
where $k^{\prime\left(\text{1D}\right)}=\hbar k/\sqrt{2mg^{\left(0\right)\left(\text{1D}\right)}n_{0}}$ denotes the dimensionless wave vector. The first two terms on the right-hand side correspond to the mean-field contribution, while the remaining terms account for beyond-mean-field corrections arising from quantum fluctuations. Notably, the final three terms in Eq.~(\ref{GSE1}) are incorporated to mitigate ultraviolet divergences through a suitable renormalization of the coupling constants, as detailed in Appendix~\ref{B}. In the continuum limit, the summation in Eq.~(\ref{GSE1}) is replaced by an integral, yielding the analytical expression for the ground-state energy of the system, given by
\begin{figure}[hbtp] 
	\begin{centering} 
		\includegraphics[scale=0.8]{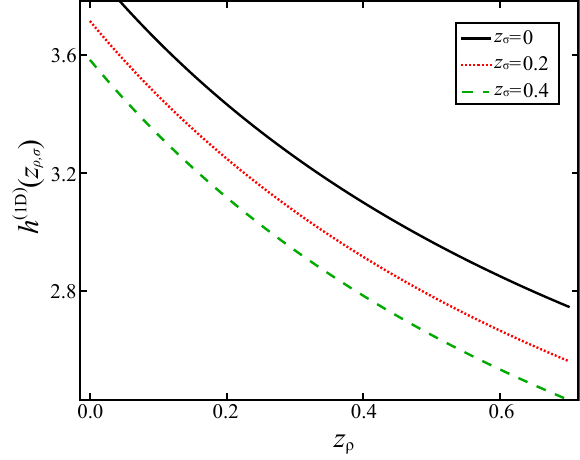} 
		\par\end{centering} 
\caption{
    Dimensionless scaling function $ h^{(\text{1D})}(z_{\rho}, z_{\sigma}) $ defined in Eq.~\eqref{GSF1}, with $ z_{\rho} \equiv 4mn_{0} g_{\rho}^{(2)}/\hbar^{2} $ continuously scanned in $[0,0.7]$ and $ z_{\sigma} \equiv 4mn_{0} g_{\sigma}^{(2)}/\hbar^{2} $ varied over discrete values in $[0,0.4]$. 
    \label{fig3}
}
\end{figure}
\begin{eqnarray}
	\frac{E_{\text{g}}^{\left(\text{1D}\right)}}{L} & = & g^{\left(0\right)\left(\text{1D}\right)}n_{0}^{2}+g^{\left(0\right)\left(\text{1D}\right)}_{12}n_{0}^{2}\nonumber \\
	&-&\frac{\left(g^{\left(0\right)\left(\text{1D}\right)}n_{0}\right)^{\frac{3}{2}}}{4\pi}\left(\frac{2m}{\hbar^{2}}\right)^{\frac{1}{2}}h^{\left(\text{1D}\right)}\left(z_{\rho,\sigma}\right),
\end{eqnarray}
where the scaling function $h_{\text{1D}}\left(z_{\rho,\sigma}\right)$ is analytically determined as
\begin{equation}
	h^{\left(\text{1D}\right)}\left(z_{\rho,\sigma}\right)= \sum_{\pm}{\frac{4\sqrt{2}}{3}\frac{\left(1\pm y^{\left(\text{1D}\right)}\right)^{\frac{3}{2}}}{\lambda_{\pm }^{\left(\text{1D}\right)}}}.
	\label{GSF1}
\end{equation}
The outcome in Eq.~(\ref{GSF1}) aligns excellently with the corresponding one in  Ref.~\cite{Chiquillo2025}. When the parameter $y^{\left(\text{1D}\right)}=0.3$, the corresponding result of Eq.~(\ref{GSF1}) are presented in Fig.~\ref{fig3}. At this point, we verify the accuracy of our results. In the limit where the finite-range interaction vanishes, the scaling function reduces to the following form
\begin{equation}
	h^{\left(0\right)\left(\text{1D}\right)}=  \sum_{\pm}{\frac{4\sqrt{2}}{3}\left(1\pm y^{\left(\text{1D}\right)}\right)^{\frac{3}{2}}},\label{GSF10}
\end{equation}
which is fully consistent with the results reported in Ref.~\cite{Chiquillo2018}. For $y^{\left(\text{1D}\right)} = 0$, the system reduces to the single-component case, and the results remain consistent with those in Ref.~\cite{Tononi2018}. In quasi-1D bosonic systems, the absence of true Bose-Einstein condensation in the thermodynamic limit, due to enhanced quantum fluctuations, precludes conventional quantum depletion, rendering this phenomenon characteristic of higher-dimensional condensates.

Eqs.~(\ref{GSE}), (\ref{QD}), and (\ref{GSE1}) constitute a pivotal result of this study, delivering explicit analytical expressions for the LHY corrections to the ground-state energy and quantum depletion influenced by finite-range interactions in 3D and quasi-1D systems, thereby elucidating the impact of effective interaction strengths on the ground-state energy and quantum depletion and providing critical insights for precisely tuning the separation of the DSF with two degrees of freedom in Sec.~\ref{4}.

Eqs.~(\ref{GSE}), (\ref{QD}), and (\ref{GSE1}) constitute a pivotal result of this study, delivering explicit analytical expressions for the LHY corrections to the ground-state energy and quantum depletion influenced by finite-range interactions in 3D and quasi-1D systems. The one-loop approximation captures leading-order quantum fluctuations, neglecting higher-order terms such as two-loop corrections, which are subdominant in the weakly interacting regime. This approach is controllable, as our results align with established limits (e.g., Ref.~\cite{Chiquillo2025} for $ z = 0 $), thereby elucidating the impact of effective interaction strengths on the ground-state energy and quantum depletion and providing critical insights for precisely tuning the separation of the DSF with two degrees of freedom in Sec.~\ref{4}.

\section{Modulating Spin-Density Separation via Finite-Range Interaction\label{4}}

In Sec.~\ref{3}, we have derived the analytical expressions for the LHY corrections to the ground-state energy and quantum depletion of 1D and 3D Bosonic mixtures, employing the one-loop approximation within the Bogoliubov framework. Building on these results, Sec.~\ref{4} aims to elucidate the role of finite-range interactions in driving spin-density separation in a Bosonic mixtures. Our analysis covers both quasi-1D and 3D regimes, contrasting dimensionality effects. Crucially, we propose an experimental protocol to probe this phenomenon through the DSF, providing a direct path to validate our predictions with experimental signatures.

\subsection{Density and spin excitation}\label{41}

In subsection~\ref{41}, our goal is to investigate how finite-range interactions can influence density and spin excitations in the system. For this purpose, we proceed to express the bosonic field $\psi_{i}$ in terms of number density $n_i$ and phase $\theta_i$ as $\psi_{i} = \sqrt{n_{i}} \mathrm{e}^{\mathrm{i}\theta_{i}} = \sqrt{n_{0} + \delta n_{i}} \mathrm{e}^{\mathrm{i}\theta_{i}}$, where $n_0$ is the equilibrium density and $\delta n_i$ denotes density fluctuations. To characterize spin-density separation, we transform to the basis $\delta n_{\rho(\sigma)} = (\delta n_1 \pm \delta n_2)/\sqrt{2}$ and $\theta_{\rho(\sigma)} = (\theta_1 \pm \theta_2)/\sqrt{2}$, corresponding to the density ($\rho$) and spin ($\sigma$) degrees of freedom. Within this framework, the Gaussian-order action in Eq.~(\ref{action}) in real time is reformulated for a $d$-dimensional system ($d = 1, 3$) as follows:
\begin{eqnarray}
	S_{\text{g}} & = & \int d^{d}\boldsymbol{r}\int dt\sum_{\gamma=\rho,\sigma}\Bigg[\frac{n_{0}\left(\nabla_{r}\theta_{\gamma}\right)^{2}}{2m}+\delta n_{\gamma}\partial_{t}\theta_{\gamma}\nonumber \\
	&  +&\frac{1}{2}\delta n_{\gamma}\left(\frac{\nabla_{r}^{2}}{4m_{\gamma}n_{0}}+g_{\gamma}^{\left(0\right)}\right)\delta n_{\gamma}\Bigg],\label{ActionGau}
\end{eqnarray}
with the coupling strength $g_{\rho\left(\sigma\right)}^{\left(0\right)}=g^{\left(0\right)}\left(1\pm y\right)$.  The reduced mass $ m_{\rho(\sigma)} = m/(1 + z_{\rho(\sigma)}) $ accounts for the effect of finite-range interactions, with the specific derivation of the finite-range action term provided in Appendix~\ref{C}. When $ z_{\rho(\sigma)} = 0 $, the effective action reduces to that of Ref.~\cite{Chung2008}. In Eq.~(\ref{ActionGau}), we emphasize that finite-range interactions, unlike other interaction types~\cite{Chung2008,Ye2025}, exclusively affect the kinetic term of density fluctuations in the Gaussian-order action, resulting in an effective mass correction. This decoupling ensures that finite-range effects modify only the number density fluctuation dynamics, without impacting the phase-related terms.

By making a Fourier transformation, the action $S_{\text{g}}$ in Eq.~(\ref{ActionGau}) can be rewritten as
\begin{equation}
	S_{g}\left(Q\right)  =  \boldsymbol{\Psi}\left(Q\right)\boldsymbol{M}^{\prime}\boldsymbol{\Psi}^{\dagger}\left(-Q\right),
\end{equation}
with $Q=\left(\boldsymbol{k},\omega\right)$ is $d+1$ vector denoting the momenta $\boldsymbol{k}$ and the frequency $\omega$, and $\boldsymbol{\Psi}\left(Q\right) =  \left(
		\delta n_{\rho},  \theta_{\rho},  \delta n_{\sigma},  \theta_{\sigma}\right)$. Furthermore, the above matrix $\boldsymbol{M}^{\prime}$ can be deduced into
\begin{equation}
	\boldsymbol{M}^{\prime}  =  \left(\begin{array}{cc}
		M_{\rho} & 0\\
		0 & M_{\sigma}
	\end{array}\right),
\end{equation}
with the $2\times2$ submatrices 
\begin{eqnarray}
	M_{\rho\left(\sigma\right)} & = & \left(\begin{array}{cc}
		\frac{\hbar^{2}k^{2}}{8m_{\rho\left(\sigma\right)}n_{0}}+\frac{g_{\rho\left(\sigma\right)}^{\left(0\right)}}{2} & \frac{1}{2}\mathrm{i}\omega_{\rho\left(\sigma\right)}\\
		-\frac{1}{2}\mathrm{i}\omega_{\rho\left(\sigma\right)} & \frac{\hbar^{2}k^{2}n_{0}}{2m}
	\end{array}\right),\label{MM}
\end{eqnarray}
being the form of two submatrices just the same as one-component inverse Green function. So we can find that unlike one-dimensional Fermi systems, which require bosonization to decouple spin and density degrees of freedom, the analysis here only necessitates considering fluctuations around the condensate density~\cite{Chung2008}. 

With the help of Eq.~(\ref{MM}), the analytical expressions of the density and spin  excitation branches can be calculated as
\begin{eqnarray}
	\varepsilon_{\rho\left(\sigma\right)}\left(k\right) & = & \sqrt{\frac{\hbar^{2}k^{2}}{2m}\left(\frac{\hbar^{2}k^{2}}{2m_{\rho\left(\sigma\right)}}+2g_{\rho\left(\sigma\right)}^{\left(0\right)}n_{0}\right)},\label{ES}
\end{eqnarray}
where $\varepsilon_{\rho}\left(k\right)$ corresponds to the density mode and $\varepsilon_{\sigma}\left(k\right)$ corresponds to the spin mode. The excitation spectrum in Eq.~(\ref{ES}) is in exact agreement with that in Eq.~(\ref{DEP}), but is derived from the perspective of spin-density separation. 

Using the excitation spectra in Eq.~\eqref{ES}, we identify contributions to the ground-state energy from density ($\rho$) and spin ($\sigma$) degrees of freedom, denoted $ E_{\rho} $ and $ E_{\sigma} $ respectively. These contributions are given by:
\begin{equation}
	\frac{E_{\rho(\sigma)}}{L^3} =  g_{\rho}^{(0)}n^2_0 + \frac{8g_{\rho(\sigma)}^{(0)\frac{5}{2}} n_{0}^{\frac{5}{2}} m^{\frac{3}{2}}}{15\pi^{2}\hbar^{3} (1 + z_{\rho(\sigma)})^{2}}.\label{DSDE}
\end{equation}
By simply adding $ E_{\rho} $ and $ E_{\sigma} $ together and noting $g^{(0)}_{\rho}=g^{(0)}+g^{(0)}_{12}$,  the obtained ground-state energy of $E_g= E_{\rho}+ E_{\sigma}$ can exactly recover the previous one in Eq.~(\ref{GSE}) as it is expected.

\subsection{Probing finite-range-induced  spin-density separation by dynamic structure factor}

We now investigate how spin-density separation in a two-component Bosonic mixtures can be experimentally probed using Bragg spectroscopy. This technique involves inducing a density perturbation in the system while simultaneously probing both spin and density excitations. By utilizing two Bragg laser beams with momenta $\boldsymbol{k}_1$ and $\boldsymbol{k}_2$ and precisely tuning their frequency difference $\omega$ (kept significantly smaller than their detuning from the atomic resonance), it is possible to selectively excite density and spin waves within the Bose gas. In what follows, we plan to calculate the DSF of finite range Bosonic mixtures in 1D and 3D cases respectively.

\subsubsection{Density and spin DSF in 3D}
To quantitatively analyze density and spin excitations, we are motivated to calculate the corresponding DSFs  directly following the method in Ref.~\cite{Chung2008},
\begin{eqnarray}
	S_{\rho\left(\sigma\right)}\left(k,\omega\right) & \approx & \frac{\chi_{\rho\left(\sigma\right)}v_{\rho\left(\sigma\right)}k\Gamma_{\rho\left(\sigma\right)}}{2\left[\left(\omega-v_{\rho\left(\sigma\right)}k\right)^{2}+\Gamma_{\rho\left(\sigma\right)}^{2}\right]}.\label{DSF3D}
\end{eqnarray}
In Eq.~(\ref{DSF3D}), the $\chi_{\rho\left(\sigma\right)}$, $v_{\rho\left(\sigma\right)}$ and $\Gamma$ are referred to as the compressibility, sound speed and damping rate respectively. In what follows, we plan to calculate these quantities one by one.

First, the compressibility $\chi_{\rho\left(\sigma\right)}$ in Eq.~(\ref{DSF3D}) can be directly calculated by ground-state energy $E$ through the expression $\chi^{-1}=\frac{1}{V}\frac{\partial^{2}E}{\partial n^{2}}$. Here, the $V$ represents the constant system volume and $n = N/V$ denotes the density. By plugging Eq.~(\ref{DSDE}) into the above definition of  compressibility, the inverse compressibility related to density ($\rho$) and spin ($\sigma$) degrees of freedom can be directly calculated as
\begin{align}
	\chi_{\rho\left(\sigma\right)}^{-1} &=  \frac{16g_{\rho\left(\sigma\right)}^{\left(0\right)\frac{5}{2}}m^{\frac{3}{2}}n_{0}^{\frac{1}{2}}z_{\rho\left(\sigma\right)}^{2}}{5\pi^{2}\hbar^{3}(z_{\rho\left(\sigma\right)}+1)^{4}}-\frac{16g_{\rho\left(\sigma\right)}^{\left(0\right)\frac{5}{2}}m^{\frac{3}{2}}n_{0}^{\frac{1}{2}}z_{\rho\left(\sigma\right)}}{3\pi^{2}\hbar^{3}(z_{\rho\left(\sigma\right)}+1)^{3}}\notag \\
	& \quad +\frac{2g_{\rho\left(\sigma\right)}^{\left(0\right)\frac{5}{2}}m^{\frac{3}{2}}n_{0}^{\frac{1}{2}}}{\pi^{2}\hbar^{3}(z_{\rho\left(\sigma\right)}+1)^{2}}+g_{\rho}^{\left(0\right)}.\label{Compress}
\end{align}

Next, the sound velocity in Eq.~(\ref{DSF3D}) is defined as $v=\left(\frac{V}{mn}\frac{\partial^{2}E}{\partial V^{2}}\right)^{1/2}$ at constant particle number $N$. By plugging Eq.~(\ref{DSDE}) into the above definition of  sound velocity, its analytical expression related to density ($\rho$) and spin ($\sigma$) degrees of freedom can be deduced as 
\begin{align}
	v_{\rho\left(\sigma\right)} &= \left[\frac{16g_{\rho\left(\sigma\right)}^{\left(0\right)\frac{5}{2}}m^{\frac{1}{2}}n_{0}^{\frac{3}{2}}z_{\rho\left(\sigma\right)}^{2}}{5\pi^{2}\hbar^{3}\left(z_{\rho\left(\sigma\right)}+1\right)^{4}}+\frac{2g_{\rho\left(\sigma\right)}^{\left(0\right)\frac{5}{2}}m^{\frac{1}{2}}n_{0}^{\frac{3}{2}}}{\pi^{2}\hbar^{3}\left(z_{\rho\left(\sigma\right)}+1\right)^{2}} \right. \notag \\
	&\quad \left. -\frac{16g_{\rho\left(\sigma\right)}^{\left(0\right)\frac{5}{2}}m^{\frac{1}{2}}n_{0}^{\frac{3}{2}}z_{\rho\left(\sigma\right)}}{3\pi^{2}\hbar^{3}\left(z_{\rho\left(\sigma\right)}+1\right)^{3}}+\frac{g_{\rho}^{\left(0\right)}n_{0}}{m}\right]^{\frac{1}{2}}.\label{Sound}
\end{align}

Before further proceeding, we would like to double-check the validity of Eqs.~(\ref{Compress}) and~(\ref{Sound}) by seeing whether they can recover the well-known results in the limit case of the vanishing finite-range interactions $z_{\rho(\sigma)} = 0$. In this limit, the compressibility in Eq.~(\ref{Compress}) simplifies to
\begin{equation}
	\chi_{\rho\left(\sigma\right)}^{\left(0\right)}  =  g_{\rho}^{\left(0\right)}+\frac{2g_{\rho\left(\sigma\right)}^{\left(0\right)\frac{5}{2}}m^{\frac{3}{2}}n_{0}^{\frac{1}{2}}}{\pi^{2}\hbar^{3}},
\end{equation}
and the sound velocity in Eq. (\ref{Sound}) simplifies to 
\begin{equation}
	v_{\rho\left(\sigma\right)}^{\left(0\right)}  =  \left[\frac{g_{\rho}^{\left(0\right)}n_{0}}{m}+\frac{2g_{\rho\left(\sigma\right)}^{\left(0\right)\frac{5}{2}}m^{\frac{1}{2}}n_{0}^{\frac{3}{2}}}{\pi^{2}\hbar^{3}}\right]^{\frac{1}{2}},
\end{equation}
consistent with the results for a contact-interaction $s$-wave Bosonic mixtures reported in Ref.~\cite{Chung2008}. These findings affirm the robustness of our theoretical framework, accurately capturing the transition from finite-range to contact interactions while maintaining the spin-density separation observed in the dynamic structure factor.

At last, to evaluate the damping rate $\Gamma$ for the DSF of a 3D two-component Bosonic mixtures with finite-range interactions, we employ the Beliaev damping framework developed for a single-component $s$-wave Bose gas, as detailed in Ref.~\cite{Chung2008}. Our calculations reveal that finite-range interactions solely influence the quadratic terms without affecting the damping mechanism. As a result, both the density and spin modes are governed by the bare mass $m$. The damping rate is thus expressed as $\Gamma=3\hbar k^{5}/(640\pi mn_{0})$, reflecting low-energy quasiparticle scattering in 3D systems, consistent with Refs.~\cite{Chung2008, Chung2009}. This $k^5$-dependent damping broadens the DSF peaks without altering the energy separation between the density and spin modes, as defined by the excitation spectra $\varepsilon_{\rho}$ and $\varepsilon_{\sigma}$ in Eq.~(\ref{ES}). Using these damping rates, we reformulate the DSFs as
\begin{eqnarray}
	S_{\rho\left(\sigma\right)} & = & \chi_{\text{\ensuremath{\rho}}}^{\left(0\right)}J_{\rho\left(\sigma\right)}\left(\omega^{\prime}\right),
\end{eqnarray}
where the 3D dimensionless DSF functions or the density and spin degrees of freedom, $J_{\rho\left(\sigma\right)}\left(\omega^{\prime}\right)$, are given by
\begin{eqnarray}
	J_{\rho\left(\sigma\right)}\left(\omega^{\prime}\right) & = & \frac{\zeta_{\rho\left(\sigma\right)}\alpha_{\rho\left(\sigma\right)}\Gamma/v_{\rho}^{\left(0\right)}k^{\prime}}{(\omega^{\prime}-\alpha_{\rho\left(\sigma\right)})^{2}+\left(\Gamma/v_{\rho}^{\left(0\right)}k^{\prime}\right)^{2}}. \label{DSF3Ddim}
\end{eqnarray}

\begin{widetext}
	
	\begin{figure}[h] 
		\begin{centering} 
			\includegraphics[scale=0.6]{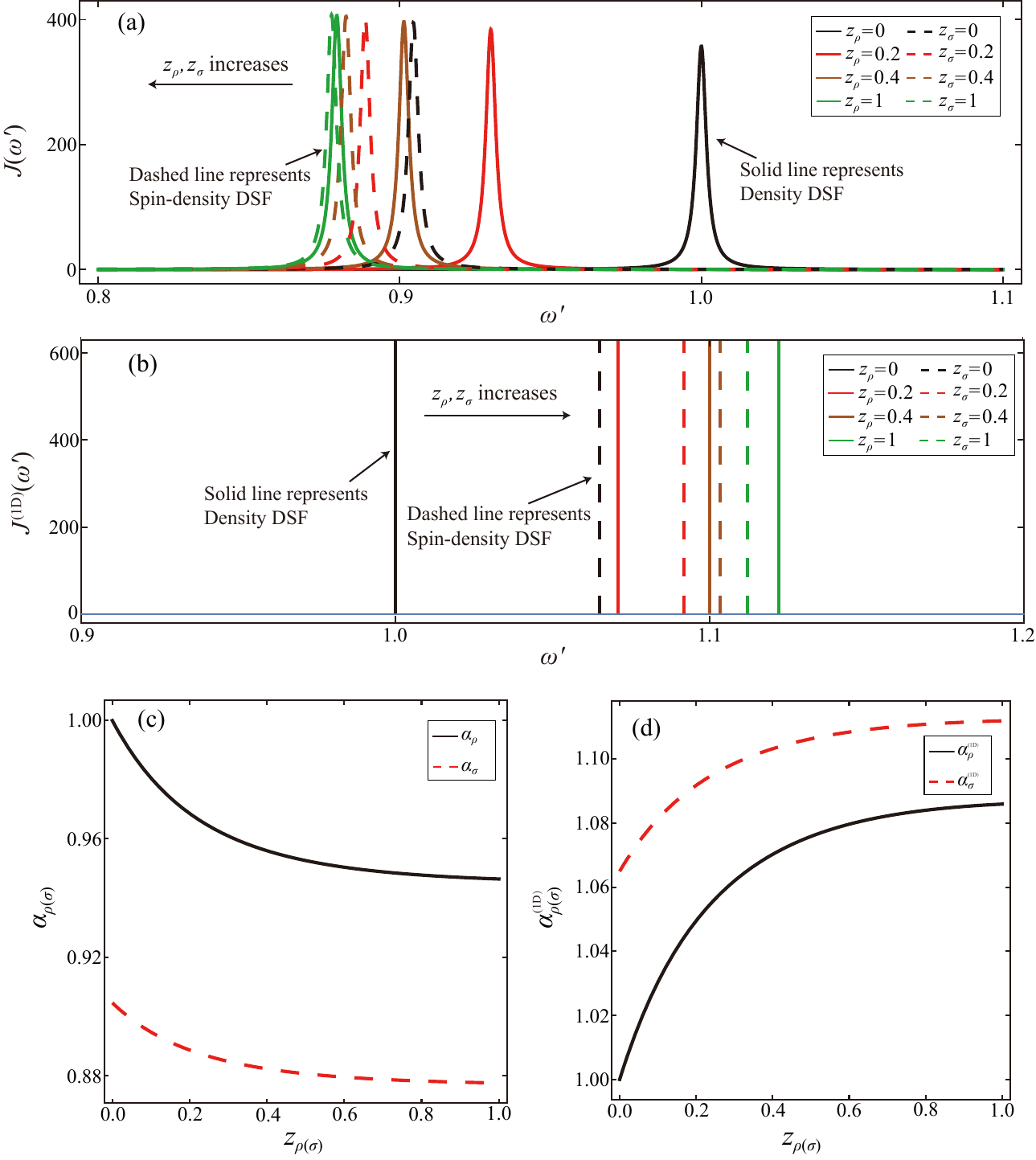} 
			\par\end{centering} 
		\caption{
    (a) 3D dimensionless DSFs $ J_\rho(\omega') $ and $ J_\sigma(\omega') $ in Eq. (\ref{DSF3Ddim}) versus $ \omega' $ for $ z_\rho, z_\sigma \in [0,1] $. Density ($\rho$) and spin ($\sigma$) peaks shift to lower frequencies with increasing finite-range interaction strength. 
    (b) 1D counterparts $ J_\rho^{\rm (1D)}(\omega') $ and $ J_\sigma^{\rm (1D)}(\omega') $ Eq. (\ref{DSF1Ddim}) versus $ \omega' $, exhibiting $\delta$-function-like peaks shifting to higher frequencies. This highlights dimensionality effects on spin-density separation. 
    (c) 3D dimensionless coefficients $ \alpha_\rho(z_\rho) $ and $ \alpha_\sigma(z_\sigma) $ versus interaction strength ($ z_{\rho,\sigma} \in [0,1] $), showing monotonic decrease. 
    (d) 1D coefficients $ \alpha_\rho^{\rm (1D)}(z_\rho) $ and $ \alpha_\sigma^{\rm (1D)}(z_\sigma) $ versus interaction strength, displaying opposite trend with monotonic increase, reflecting dimension-dependent separation dynamics.
    \label{figf}
}
	\end{figure}
	
\end{widetext}

In Eq.~(\ref{DSF3Ddim}), the dimensionless coefficients are defined as $\alpha_{\rho\left(\sigma\right)}=v_{\rho\left(\sigma\right)}/v_{\rho,\text{3D}}^{\left(0\right)}$ and $\zeta_{\rho\left(\sigma\right)}=\chi_{\rho}^{\left(0\right)-1}/\chi_{\rho\left(\sigma\right)}^{-1}$, with the dependence of $\alpha_{\rho(\sigma)}$ on the interaction parameters $z_{\rho(\sigma)}$ depicted in Fig.~\ref{figf}(c). The momentum $k$ is rendered dimensionless, as outlined in Sec.~\ref{2}, where $k^{\prime} = \hbar k / \sqrt{2g n_0 m}$, with $\hbar / \sqrt{2g n_0 m} = 1$. Consequently, the dimensionless frequency is given by $\omega^{\prime}=\omega/v_{\rho}^{\left(0\right)}k^{\prime}$. For $k^{\prime} = 1.25$ and $y = 0.3$, the density compressibility $\chi_{\text{\ensuremath{\rho}}}^{\left(0\right)}$ can be approximated as constant. The dimensionless DSF functions $J_{\rho(\sigma)}(\omega^{\prime})$ vary with frequency $\omega^{\prime}$ for different values of the interaction parameters $z_{\rho}$ and $z_{\sigma}$, as illustrated in Fig.~\ref{figf}(a). From Fig.~\ref{figf}(a), we observe that the DSF peaks for both density and spin modes shift toward lower frequencies with increasing finite-range interaction strength, indicating enhanced interaction effects. The density wave exhibits a broader peak due to stronger Beliaev damping, whereas the spin wave retains a narrower and sharper response, underscoring their distinct dynamic behaviors in 3D Bosonic mixtures.

\subsubsection{Density and spin DSF in 1D}
Having analyzed spin-density separation in 3D Bosonic mixtures through dynamic structure factors, we now investigate the quasi-1D case to further elucidate this phenomenon. 
Following Ref.~\cite{Chung2008},  the formal expression of the DSFs in 1D is given by
\begin{eqnarray}
	S_{\rho(\sigma)}^{\text{1D}}\left(k,\omega\right)  =  \text{Im}\frac{\chi_{\rho\left(\sigma\right)}^{\left(\text{1D}\right)}v_{\rho\left(\sigma\right)}^{\left(\text{1D}\right)2}k^{2}}{\omega^{2}-\omega_{\rho\left(\sigma\right)}^{2}\left(k\right)},\label{DSF1D}
\end{eqnarray}
with  $\chi_{\rho(\sigma),\text{1D}}$ representing the compressibility, $v_{\rho\left(\sigma\right)}^{\left(\text{1D}\right)}$ being the sound velocity and $\omega_{\rho\left(\sigma\right)}$ labelling the density and spin excitations. In what follows, we plan to obtain the analytical expressions of these three quantities.

As a first step, the compressibility of $\chi_{\rho(\sigma),\text{1D}}$ is determined by $\chi^{-1} = \frac{1}{L} \frac{\partial^2 E}{\partial n^2}$ at constant system length $L$. We need to calculate the contribution to the ground-state energy of the density and spin excitations provided in Eq.~(\ref{ES}), which are given as
\begin{equation}
	\frac{E_{\rho\left(\sigma\right)}^{\left(\text{1D}\right)}}{L} \! = \! \frac{g_{\rho}^{\left(0\right)\left(\text{1D}\right)}n_{0}^{2}}{2}-\frac{2}{3\pi}\left(\frac{m}{\hbar^{2}}\right)^{\frac{1}{2}}\frac{\left(g_{\rho\left(\sigma\right)}^{\left(0\right)\left(\text{1D}\right)}n_{0}\right)^{\frac{3}{2}}}{\left(1+z_{\rho\left(\sigma\right)}\right)}.
\end{equation}
The above results agrees with the corresponding ones in Ref.~\cite{cappel2017} and in the limit $z_{\rho(\sigma)} = 0$ can reduce to well-known Lieb-Liniger solution~\cite{Lieb1963}.
We proceed to calculate the inverse compressibility  $\chi^{-1} $ induced by the density and spin excitations as
\begin{align}
	\chi_{\rho\left(\sigma\right)}^{\left(\text{1D}\right)-1} &=  -\frac{4g_{\rho\left(\sigma\right)}^{\left(0\right)\left(\text{1D}\right)\frac{3}{2}}m^{\frac{1}{2}}z_{\rho\left(\sigma\right)}^{2}}{3\pi n_{0}^{\frac{1}{2}}\left(z_{\rho\left(\sigma\right)}+1\right)^{3}}+\frac{2g_{\rho\left(\sigma\right)}^{\left(0\right)\left(\text{1D}\right)\frac{3}{2}}m^{\frac{1}{2}}z_{\rho\left(\sigma\right)}}{\pi n_{0}^{\frac{1}{2}}\left(z_{\rho\left(\sigma\right)}+1\right)}\notag\\
	&\quad -\frac{g_{\rho\left(\sigma\right)}^{\left(0\right)\left(\text{1D}\right)\frac{3}{2}}m^{\frac{1}{2}}}{2\pi n_{0}^{\frac{1}{2}}\left(z_{\rho\left(\sigma\right)}+1\right)}+g_{\rho}^{\left(0\right)\left(\text{1D}\right)}.\label{Compress1D}
\end{align}

Next, we are routine to obtain the analytical expression of the quasi-1D sound velocity of $v_{\rho\left(\sigma\right)}^{\left(\text{1D}\right)}$ in Eq.~(\ref{DSF1D}), reading
\begin{align}
	\nu_{\rho\left(\sigma\right)}^{\left(\text{1D}\right)} &= \left[-\frac{4g_{\rho\left(\sigma\right)}^{\left(0\right)\left(\text{1D}\right)\frac{3}{2}}n_{0}^{\frac{1}{2}}z_{\rho\left(\sigma\right)}^{2}}{3\pi m^{\frac{1}{2}}\left(z_{\rho\left(\sigma\right)}+1\right)^{3}}+\frac{2g_{\rho\left(\sigma\right)}^{\left(0\right)\left(\text{1D}\right)\frac{3}{2}}n_{0}^{\frac{1}{2}}z_{\rho\left(\sigma\right)}}{\pi m^{\frac{1}{2}}\left(z_{\rho\left(\sigma\right)}+1\right)^{2}}\right. \notag\\
	&\quad \left. -\frac{g_{\rho\left(\sigma\right)}^{\left(0\right)\left(\text{1D}\right)\frac{3}{2}}n_{0}^{\frac{1}{2}}}{2\pi m^{\frac{1}{2}}\left(z_{\rho\left(\sigma\right)}+1\right)}+\frac{g_{\rho}^{\left(0\right)\left(\text{1D}\right)}n_{0}}{m}\right]^{\frac{1}{2}}.\label{Sound1D}
\end{align}

At last, we remark that in quasi-1D systems, the absence of Beliaev damping results in sharp DSF peaks, as quasiparticle decay processes are suppressed due to kinematic constraints. The DSFs in Eq.~(\ref{DSF1D})  can be further deduced into 
\begin{eqnarray}
	S_{\rho(\sigma)}^{\left(\text{1D}\right)}\left(k,\omega\right) & \approx & \frac{\pi\chi_{\rho\left(\sigma\right)}^{\left(\text{1D}\right)}v_{\rho\left(\sigma\right)}^{\left(\text{1D}\right)}k}{2}\delta\left(\omega-v_{\rho\left(\sigma\right)}^{\left(\text{1D}\right)}k\right),\nonumber \\
	\label{DSF1D1}
\end{eqnarray}
reflecting a $\delta$-function-like response in quasi-1D systems. The detailed derivation from Eq.~(\ref{DSF1D}) to Eq.~(\ref{DSF1D1}) is provided in Appendix~\ref{C}. By plugging Eq.~(\ref{Compress1D}) and (\ref{Sound1D}) into Eq.~(\ref{DSF1D1}), we succeed to obtain the analytical expression of the DSFs for a Bosonic mixtures with the finite range interaction in 1D.

To facilitate the analysis of spin-density separation in quasi-1D systems, we perform a dimensionless treatment of the DSFs, reformulating them as
\begin{eqnarray}
	S_{\rho\left(\sigma\right)}^{\left(\text{1D}\right)}\left(k,\omega\right) & \approx & \chi_{\rho}^{\left(0\right)\left(\text{1D}\right)}J_{\rho\left(\sigma\right)}^{\left(\text{1D}\right)}\left(\omega^{\prime}\right),\label{DSF1D11}
\end{eqnarray}
with $\chi_{\rho}^{\left(0\right)\left(\text{1D}\right)-1}  =  -g_{\rho}^{\left(0\right)\left(\text{1D}\right)3/2}m^{1/2}/2\pi n_{0}^{1/2}+g_{\rho}^{\left(0\right)\left(\text{1D}\right)}$ and $\omega^{\prime}=\omega/v_{\rho}^{(0)\text{1D}}k^{\prime(\text{1D})}$.
 The quasi-1D dimensionless DSFs function in Eq.~(\ref{DSF1D11}) are given by
\begin{equation}
	J_{\rho\left(\sigma\right)}^{\left(\text{1D}\right)}\left(\omega^{\prime}\right)  =  \frac{\pi\chi_{\rho}^{\left(0\right)\left(\text{1D}\right)-1}\alpha_{\rho\left(\sigma\right)}^{\left(\text{1D}\right)}}{2\chi_{\rho\left(\sigma\right)}^{\left(\text{1D}\right)-1}}\delta\left(\omega^{\prime}-\alpha_{\rho\left(\sigma\right)}^{\left(\text{1D}\right)}\right).\label{DSF1Ddim}
\end{equation}

Now, we are ready to investigate how the finite-range interaction can affect the spin-density separation in 1D by Eq.~(\ref{DSF1Ddim}). 
The quasi-1D dimensionless coefficient in Eq.~(\ref{DSF1Ddim}) is defined as $\alpha_{\rho\left(\sigma\right)}^{\left(\text{1D}\right)}=v_{\rho\left(\sigma\right)}^{\left(\text{1D}\right)}/v_{\rho}^{\left(0\right)\left(\text{1D}\right)}$, which we illustrate in Fig.~\ref{figf}(d) by plotting their variation with the interaction parameters $z_{\rho(\sigma)}^{\left(\text{1D}\right)}$. With the dimensionless frequency defined as $\omega^{\prime}=\omega/\left(v_{\rho}^{(0)\left(\text{1D}\right)}k^{\prime\left(\text{1D}\right)}\right)$, we analyze the 1D DSFs to characterize spin-density separation. Setting $k^{\prime\left(\text{1D}\right)}=1.25$ and $y=0.3$, the density compressibility $\chi_{\rho}^{(0)\left(\text{1D}\right)}$ is approximated as constant. The dimensionless DSF functions $J_{\rho(\sigma)}^{\left(\text{1D}\right)}(\omega^{\prime})$ vary with $\omega^{\prime}$ for different values of $z_{\rho}$ and $z_{\sigma}$, as illustrated in Fig.~\ref{figf}(b). From Fig.~\ref{figf}(b), the DSF peaks for both density and spin modes, exhibiting $\delta$-function-like behavior, shift to higher frequencies with increasing finite-range interaction strength, in contrast to the 3D case where peaks shift to lower frequencies. This distinction highlights the unique dynamic signatures of spin-density separation across dimensions.

\section{CONCLUSION\label{5}}

This work establishes the ground-state energy and quantum depletion with finite-range interactions and elucidates their role in governing spin-density separation within Bosonic mixtures. In Sec.~\ref{3}, we compute the ground-state energy and quantum depletion for 3D and 1D systems, including LHY corrections to the ground-state energy and quantum depletion. Section~\ref{4} visually demonstrates spin-density separation through DSFs, revealing that finite-range interactions—parametrized by $ z_{\rho(\sigma)} $—enable continuous tuning of spectral peak positions in both spin ($\sigma$) and density ($\rho$) channels. This highlights a dimension-dependent inversion of peak shifts: in 1D systems, DSF peaks shift to higher frequencies with increasing interaction strength, whereas in 3D they shift to lower frequencies, underscoring dimensionality's critical role in separation dynamics.

Our theoretical framework is validated through dimension-specific \textit{a posteriori} checks:
\begin{itemize}
    \item \textbf{3D verification}: Quantum depletion $\frac{N - N_0}{N}$ (Eq.~\eqref{QD}) is evaluated as $\approx 0.0028 \times I(z_{\rho(\sigma)})$ using experimental parameters $n \approx 3 \times 10^{14}~\text{cm}^{-3}$, $a \approx 2.7517~\text{nm}$ (Fig.~\ref{fig2}(b)). The resulting $\mathcal{O}(10^{-3})$ depletion confirms Bogoliubov approximation validity~\cite{Xu2006,Knoop2011}.
    
    \item \textbf{Quasi-1D verification}: Ground-state energy calculations via coherent-state path integrals employ a continuous momentum approximation, justified by $L \gg a_{\perp} = \sqrt{\hbar / m \omega_{\perp}}$ and low-energy excitations ($k a_{\perp} \ll 1$). This is consistent with optical-trap experiments where $L/a_{\perp} \sim 10^3$ ensures homogeneity in the central region~\cite{Jacqmin2012,Fabbri2011}.
    
    \item \textbf{Parameter validation}: Effective ranges $r_{\mathrm{e}}$ tunable via dark-state control ($-1.4 \times 10^6$ to $4.2 \times 10^6~a_0$, where $a_0$ is the Bohr radius)~\cite{Wu2012} yield $z_{\rho(\sigma)} \gg 1$, confirming our $[0,1]$ range covers physical regimes.
\end{itemize}

The experimental realization of our proposal requires precise control over four key parameters: the inter- and intraspecies $s$-wave scattering lengths $ a_{12} $ and $ a $, and their corresponding effective ranges $ r_{\mathrm{e}_{12}} $ and $ r_{\mathrm{e}} $. These parameters are highly tunable in both 3D and quasi-1D systems via identical state-of-the-art techniques: Feshbach resonances for scattering lengths~\cite{Inouye1998,Tanzi2018,Jacqmin2012} and dark-state control for effective ranges~\cite{Wu2012}. By engineering these parameters, our predictions for spin-density separation can be probed through Bragg spectroscopy measurements of the dynamic structure factor $ S(k, \omega) $—a technique already implemented in both dimensional regimes~\cite{Stenger1999,Richard2003}. We anticipate that the distinctive influence of finite-range interactions on spin-density separation will be validated in future experiments across these dimensional settings.  

In summary, this work explores the impact of finite-range interactions on the ground-state energy and quantum depletion and spin-density separation in Bosonic mixtures across 3D and quasi-1D regimes. At the Gaussian level, we establish that finite-range interactions decouple spin degrees of freedom, modifying the effective mass as analytically derived in Sec.~\ref{4}. This non-local-interaction-driven decoupling manifests in distinct DSF peak shifts: to lower frequencies in 3D and higher frequencies in quasi-1D. By tuning the range parameters $ r_{\mathrm{e}} $ and $ r_{\mathrm{e}_{12}} $, the DSF peaks of density ($\rho$) and spin ($\sigma$) modes become continuously adjustable, enabling precise control of spin-density separation and significantly enriching dynamical behavior beyond contact-interaction limits. While we approximate finite-range strengths using the 1D coupling $ g^{(2)} $ from Refs.~\cite{Lorenzi2023,cappel2017}, deriving rigorously dimensionally-reduced interaction strengths for quasi-1D systems remains an open challenge. Incorporating second-order couplings $ g^{(2)} $ from 3D reduction could enhance DSF prediction accuracy and experimental validation. By bridging theoretical insights with experimental accessibility through Bragg spectroscopy protocols, this work advances understanding of interaction-driven phenomena in quantum gases and establishes a foundation for exploring dimensionally-tuned exotic phases governed by finite-range interactions.

\section*{ACKNOWLEDGMENTS}
We thank Tao Yu, Ying Hu, and Biao Wu for stimulating discussions and useful help. This work was supported by the National Natural Science Foundation of China (Grants No. 12574301) and  the Zhejiang Provincial Natural Science Foundation (Grant No. LZ25A040004).

\appendix
\begin{widetext}
\section{CAYLEY-HAMILTON THEOREM\label{A}}

From the Cayley-Hamilton theorem, we can get the excitation (\ref{DEP})
with Eq.~(\ref{m}), the matrix $\boldsymbol{M}$, which reads
\begin{eqnarray}
	\boldsymbol{M} & = & \left(\begin{array}{cccc}
		\frac{\hbar^{2}k^{2}}{2m}+g^{\prime}n_{0} &g^{\prime}n_{0} & g^{\prime}_{12}n_{0} & g^{\prime}_{12}n_{0}\\
	g^{\prime}n_{0} & \frac{\hbar^{2}k^{2}}{2m}+g^{\prime}n_{0} & g^{\prime}_{12}n_{0} & g^{\prime}_{12}n_{0}\\
	g^{\prime}_{12}n_{0} & g^{\prime}_{12}n_{0} & \frac{\hbar^{2}k^{2}}{2m}+g^{\prime}n_{0} & g^{\prime}n_{0}\\
		g^{\prime}_{12}n_{0} & g^{\prime}_{12}n_{0} & g^{\prime}n_{0} & \frac{\hbar^{2}k^{2}}{2m}+g^{\prime}n_{0}
	\end{array}\right),
\end{eqnarray}
where $g^{\prime}=g^{\left(0\right)}+g^{\left(2\right)}k^{2}$, $g^{\prime}_{12}=g^{\left(0\right)}_{12}+g^{\left(2\right)}_{12}k^{2}$,
and $\boldsymbol{\kappa}=\left(\begin{array}{cccc}
	1 & 0 & 0 & 0\\
	0 & -1 & 0 & 0\\
	0 & 0 & 1 & 0\\
	0 & 0 & 0 & -1
\end{array}\right)$. Our starting point is the trace term in Eq.~(\ref{DE0}), which
takes form
\begin{eqnarray}
	\frac{1}{4}\text{Tr}\left[\left(\boldsymbol{\kappa M}\right)^{2}\right] & = & \frac{\hbar^{2}k^{2}}{2m}\left(\frac{\hbar^{2}k^{2}}{2m}+2g^{\prime}n_{0}\right).
\end{eqnarray}

By nondimensionalizing it, we obtain
\begin{eqnarray*}
	\frac{1}{4}\text{Tr}\left[\left(\boldsymbol{\kappa M}\right)^{2}\right] & = & \frac{\hbar^{2}k^{2}}{2m}\left(\frac{\hbar^{2}k^{2}}{2m}+2g^{\prime}n_{0}\right)\\
	& = & \frac{\hbar^{2}k^{2}}{2m}\left(\frac{\hbar^{2}k^{2}}{2m}+2g^{\left(0\right)}n_{0}+2g^{\left(2\right)}n_{0}k^{2}\right),
\end{eqnarray*}
then we can calculate the determinant term, which reads
\begin{eqnarray}
	\det\left(\boldsymbol{\kappa M}\right) & = & \left(\frac{\hbar^{2}k^{2}}{2m}\right)^{2}\left(\frac{\hbar^{2}k^{2}}{2m}+2n_{0}\left(g^{\prime}-g^{\prime}_{12}\right)\right)\left(\frac{\hbar^{2}k^{2}}{2m}+2n_{0}\left(g^{\prime}+g^{\prime}_{12}\right)\right).
\end{eqnarray}

So Eq.~(\ref{DE0}) can be rewritten as

\begin{eqnarray}
	E_{\pm} & = & \sqrt{\frac{1}{4}{\rm Tr}\left(\left(\boldsymbol{\kappa}\boldsymbol{M}\right)^{2}\right)\pm\sqrt{\frac{1}{16}\left[{\rm Tr}\left(\boldsymbol{\kappa}\boldsymbol{M}\right)^{2}\right]^{2}-\det\left(\boldsymbol{\kappa}\boldsymbol{M}\right)}}\label{AppE}\nonumber \\
	& = & \sqrt{\frac{\hbar^{2}k^{2}}{2m}\left(\frac{\hbar^{2}k^{2}}{2m}+2g^{\prime}n_{0}\right)\pm\sqrt{\left(\frac{\hbar^{2}k^{2}}{2m}\right)^{2}\left[\left(\frac{\hbar^{2}k^{2}}{2m}+2g^{\prime}n_{0}\right)^{2}-\left(\left(\frac{\hbar^{2}k^{2}}{2m}+2n_{0}g^{\prime}\right)^{2}-\left(2n_{0}g^{\prime}_{12}\right)^{2}\right)\right]}}\nonumber\\
	& = & \sqrt{\frac{\hbar^{2}k^{2}}{2m}\left(\frac{\hbar^{2}k^{2}}{2m}+2g^{\left(0\right)}n_{0}\pm2g^{\left(0\right)}_{12}n_{0}+\frac{\hbar^{2}k^{2}}{2m}\left[\frac{4mn_{0}\left(g^{\left(2\right)}\pm g^{\left(2\right)}_{12}\right)}{\hbar^{2}}\right]\right)}\nonumber\\
	& = & \sqrt{\frac{\hbar^{2}k^{2}}{2m}\left(\frac{\hbar^{2}k^{2}}{2m}\lambda_{\pm}+2g^{\left(0\right)}n_{0}\pm2g^{\left(0\right)}_{12}n_{0}\right)}\nonumber\\
	& = & g^{\left(0\right)}n_{0}\sqrt{\frac{\hbar^{2}k^{2}}{2mg^{\left(0\right)}n_{0}}\left(\frac{\hbar^{2}k^{2}}{2mg^{\left(0\right)}n_{0}}\lambda_{\pm}+2\pm2y\right)},
\end{eqnarray}
which is exactly Eq.~(\ref{DEP}).

\section{REMOVE POWER ULTRAVIOLET DIVERGENCE\label{B}}

In this appendix, we follow Ref.~\cite{Braaten1997} to derive the crucial regularizing
terms on the second line of Eq.~(\ref{GSE3}) and Eq.~(\ref{GSE1}), take 3D case as example, allowing the ground-state energy to be expressed as

\begin{eqnarray}
	\frac{E_{\text{g}}}{V} & = & g^{\left(0\right)}n_{0}^{2}+g^{\left(0\right)}_{12}n_{0}^{2}+\frac{g^{\left(0\right)}n_{0}}{2V}\sum_{k\ne0}\left\{ f_{+}+f_{-}-\lim_{k\rightarrow\infty}\left(f_{-}+f_{+}\right)\right\} ,\label{OGS}
\end{eqnarray}
with 
\begin{eqnarray}
	\lim_{k\rightarrow\infty}f_{+} & = & \lim_{k\rightarrow\infty}\sqrt{\frac{{\rm Tr}\left(\left(\boldsymbol{\kappa}\boldsymbol{M}\right)^{2}\right)}{4\left(g^{\left(0\right)}n_{0}\right)^{2}}+\sqrt{\frac{\left[{\rm Tr}\left(\boldsymbol{\kappa}\boldsymbol{M}\right)^{2}\right]^{2}}{16\left(g^{\left(0\right)}n_{0}\right)^{4}}-\frac{\det\left(\boldsymbol{\kappa}\boldsymbol{M}\right)}{\left(g^{\left(0\right)}n_{0}\right)^{4}}}}\nonumber \\
	& = & -\frac{\left(1+y\right)^{2}}{2\lambda_{+}^{\frac{3}{2}}\frac{\hbar k^{2}}{2mg^{\left(0\right)}n_{0}}}+\frac{1+y}{\sqrt{\lambda_{+}}}+\frac{\hbar k^{2}}{2mg^{\left(0\right)}n_{0}}\sqrt{\lambda_{+}},
\end{eqnarray}
and
\begin{eqnarray}
	\lim_{k\rightarrow\infty}f_{-} & = & \lim_{k\rightarrow\infty}\sqrt{\frac{{\rm Tr}\left(\left(\boldsymbol{\kappa}\boldsymbol{M}\right)^{2}\right)}{4\left(g^{\left(0\right)}n_{0}\right)^{2}}-\sqrt{\frac{\left[{\rm Tr}\left(\boldsymbol{\kappa}\boldsymbol{M}\right)^{2}\right]^{2}}{16\left(g^{\left(0\right)}n_{0}\right)^{4}}-\frac{\det\left(\boldsymbol{\kappa}\boldsymbol{M}\right)}{\left(g^{\left(0\right)}n_{0}\right)^{4}}}}\nonumber \\
	& = & \frac{\hbar k^{2}}{2mg^{\left(0\right)}n_{0}}\sqrt{\lambda_{-}}+\frac{1+y}{\sqrt{\lambda_{-}}}-\frac{\left(1+y\right)^{2}}{2\lambda_{-}^{\frac{3}{2}}\frac{\hbar k^{2}}{2mg^{\left(0\right)}n_{0}}},
\end{eqnarray}
So we get the ultraviolet divergence in Eq.~(\ref{GSE3}) and Eq.~(\ref{GSE1}).

\section{THE HYDRODYNAMIC ACTION OF THE FINITE-RANGE TERM\label{C}}
Considering density fluctuations, we can re-express the Lagrangian density of the terms related to finite-range interactions in the action Eq.~(\ref{action}) as:
\begin{eqnarray}
	\mathcal{L}_{\text{finite}} & = & -\frac{g_{2}}{2}\left(n_{0}+\delta n_{1}\right)\nabla^{2}\left(n_{0}+\delta n_{1}\right)-\frac{g_{2}}{2}\left(n_{0}+\delta n_{2}\right)\nabla^{2}\left(n_{0}+\delta n_{2}\right)\nonumber \\
	&  & -\frac{g_{212}}{2}\left(\left(n_{0}+\delta n_{1}\right)\nabla^{2}\left(n_{0}+\delta n_{2}\right)+\left(n_{0}+\delta n_{2}\right)\nabla^{2}\left(n_{0}+\delta n_{1}\right)\right)\nonumber\\
	& = & -\frac{g_{2}}{4}\left(\delta n_{\rho}\nabla^{2}\delta n_{\rho}+\delta n_{\rho}\nabla^{2}\delta n_{\sigma}+\delta n_{\sigma}\nabla^{2}\delta n_{\rho}+\delta n_{\sigma}\nabla^{2}\delta n_{\sigma}\right)\nonumber\\
	&  & -\frac{g_{2}}{4}\left(\delta n_{\rho}\nabla^{2}\delta n_{\rho}-\delta n_{\rho}\nabla^{2}\delta n_{\sigma}-\delta n_{\sigma}\nabla^{2}\delta n_{\rho}+\delta n_{\sigma}\nabla^{2}\delta n_{\sigma}\right)\nonumber \\
	&  & -\frac{g_{212}}{4}\left(\delta n_{\rho}\nabla^{2}\delta n_{\rho}-\delta n_{\rho}\nabla^{2}\delta n_{\sigma}+\delta n_{\sigma}\nabla^{2}\delta n_{\rho}-\delta n_{\sigma}\nabla^{2}\delta n_{\sigma}\right)\nonumber \\
	&  & -\frac{g_{212}}{4}\left(\delta n_{\rho}\nabla^{2}\delta n_{\rho}+\delta n_{\rho}\nabla^{2}\delta n_{\sigma}-\delta n_{\sigma}\nabla^{2}\delta n_{\rho}-\delta n_{\sigma}\nabla^{2}\delta n_{\sigma}\right)\nonumber \\
	& = & -\frac{\left(g_{2}+g_{212}\right)}{2}\delta n_{\rho}\nabla^{2}\delta n_{\rho}-\frac{\left(g_{2}-g_{212}\right)}{2}\delta n_{\sigma}\nabla^{2}\delta n_{\sigma}\nonumber\\
	& = & -\frac{g_{2\rho}}{2}\delta n_{\rho}\nabla^{2}\delta n_{\rho}-\frac{g_{2\sigma}}{2}\delta n_{\sigma}\nabla^{2}\delta n_{\sigma}.
\end{eqnarray}
Thus, the action can be obtained as:
\begin{eqnarray}
	S&=&\frac{g_{2\rho}}{2}k^{2}\delta n_{\rho}\left(k,\omega\right)\delta n_{\rho}\left(-k,-\omega\right)+\frac{g_{2\sigma}}{2}k^{2}\delta n_{\sigma}\left(k,\omega\right)\delta n_{\sigma}\left(-k,-\omega\right).
\end{eqnarray}

\section{DETAILED DERIVATION OF EQ.~(\ref{DSF1D1}) \label{D}}

The purpose of this Appendix is to give a detailed derivation of Eq.~(\ref{DSF1D1}).
Starting from Eq.~(\ref{DSF1D}), we can get

\begin{eqnarray}
	S_{\rho(\sigma)}^{\left(\text{1D}\right)}\left(k,\omega\right) & = & \text{Im}\frac{\chi_{\rho\left(\sigma\right)}^{\left(\text{1D}\right)}v_{\rho\left(\sigma\right)}^{\left(\text{1D}\right)2}k^{2}}{\omega^{2}-\omega_{\rho\left(\sigma\right)}^{\left(\text{1D}\right)2}\left(k\right)}\nonumber\\
	& = & \text{Im}\chi_{\rho\left(\sigma\right)}^{\left(\text{1D}\right)}v_{\rho\left(\sigma\right)}^{\left(\text{1D}\right)2}k^{2}\frac{1}{2v_{\rho,\sigma}^{\left(\text{1D}\right)}k}\left(\frac{1}{\omega-v_{\rho\left(\sigma\right)}^{\left(\text{1D}\right)}k}-\frac{1}{\omega+v_{\rho\left(\sigma\right)}^{\left(\text{1D}\right)}k}\right)\nonumber\\
	& \overset{\gamma\rightarrow0^{+}}{\approx} & \text{Im}\frac{\chi_{\rho\left(\sigma\right)}^{\left(\text{1D}\right)}v_{\rho\left(\sigma\right)}^{\left(\text{1D}\right)}k}{2}\left(\frac{1}{\omega-v_{\rho\left(\sigma\right)}^{\left(\text{1D}\right)}k-\mathrm{i}\gamma}-\frac{1}{\omega+v_{\rho\left(\sigma\right)}^{\left(\text{1D}\right)}k-\mathrm{i}\gamma}\right)\nonumber\\
	& \overset{\gamma\rightarrow0^{+}}{=} & \frac{\pi\chi_{\rho\left(\sigma\right)}^{\left(\text{1D}\right)}v_{\rho\left(\sigma\right)}^{\left(\text{1D}\right)}k}{2}\left[\delta\left(\omega-v_{\rho\left(\sigma\right)}^{\left(\text{1D}\right)}k\right)-\delta\left(\omega+v_{\rho\left(\sigma\right)}^{\left(\text{1D}\right)}k\right)\right]\nonumber\\
	& \overset{\omega>0}{=} & \frac{\pi\chi_{\rho\left(\sigma\right)}^{\left(\text{1D}\right)}v_{\rho\left(\sigma\right)}^{\left(\text{1D}\right)}k}{2}\delta\left(\omega-v_{\rho\left(\sigma\right)}^{\left(\text{1D}\right)}k\right).\label{DSFA}
\end{eqnarray}
So we get the Eq.~(\ref{DSF1D1}).
\end{widetext}

\bibliography{xyref}
\end{document}